\documentclass{article}

\usepackage[nonatbib,preprint]{neurips_2022}

\usepackage[utf8]{inputenc} 
\usepackage[T1]{fontenc}    
\usepackage{hyperref}       
\usepackage{url}            
\usepackage{booktabs}       
\usepackage{amsfonts}       
\usepackage{nicefrac}       
\usepackage{microtype}      
\usepackage{xcolor}         
\usepackage{bbm}
\usepackage{bm}
\usepackage{amsmath}
\usepackage{graphicx}
\usepackage{fancyhdr}
\usepackage{bbold}
\newcommand{\tr}{^{\top}}

\title{A probabilistic framework for task-aligned intra- and inter-area neural manifold estimation}

\author{
  Edoardo Balzani\\
  Center for Neural Science\\
  New York University\\
  New York, NY, 10003 \\
  \texttt{eb162@nyu.edu} \\
  \And
  Jean Paul Noel\\
  Center for Neural Science\\
  New York University\\
  New York, NY, 10003 \\
  \texttt{jpn5@nyu.edu}
  \And
  Pedro Herrero-Vidal\\
  Center for Neural Science\\
  New York University\\
  New York, NY, 10003 \\
  \texttt{pmh314@nyu.edu}
  \AND
  Dora E. Angelaki\\
  Center for Neural Science\\
  New York University\\
  New York, NY, 10003 \\
  \texttt{da93@nyu.edu}
  \And
  Cristina Savin\\
  Center for Neural Science\\
  Center for Data Science\\
  New York University\\
  New York, NY, 10003\\
  \texttt{cs5360@nyu.edu}
}

\begin{document}

\maketitle

\begin{abstract}
  
Latent manifolds provide a compact characterization of neural population activity and of shared co-variability across brain areas. Nonetheless, existing statistical tools for extracting neural manifolds face limitations in terms of interpretability of latents with respect to task variables, and can be hard to apply to datasets with no trial repeats. Here we propose a novel probabilistic framework that allows for interpretable partitioning of population variability within and across areas in the context of naturalistic behavior. Our approach for task aligned manifold estimation (TAME-GP) extends a probabilistic variant of demixed PCA by (1) explicitly partitioning variability into private and shared sources, (2) using a Poisson noise model, and (3) introducing temporal smoothing of latent trajectories in the form of a Gaussian Process prior. This TAME-GP graphical model allows for robust estimation of task-relevant variability in local population responses, and of shared co-variability between brain areas. We demonstrate the efficiency of our estimator on within model and biologically motivated simulated data. We also apply it to neural recordings in a closed-loop virtual navigation task in monkeys, demonstrating the capacity of TAME-GP to capture meaningful intra- and inter-area neural variability with single trial resolution.
\end{abstract}

\section{Introduction}
Systems neuroscience is gradually shifting from relatively simple and controlled tasks, to studying naturalistic closed-loop behaviors where no two observations (i.e., "trials") are alike \cite{michaiel2020dynamics,shamash2021mice,noel2021flexible}. Concurrently, neurophysiological techniques are advancing rapidly \cite{stevenson2011advances,jun2017fully,angotzi2019sinaps,boi2020multi,steinmetz2021neuropixels} allowing researchers to record from an ever-increasing number of simultaneous neurons (i.e., "neural populations") and across multiple brain areas. These trends lead to a pressing need for statistical tools that compactly characterize the statistics of neural activity within and across brain regions. Dimensionality reduction techniques (reviewed below) are a popular tool for interrogating the structure of neural responses \cite{cunningham2014dimensionality}. However, as neural responses are driven by increasingly complex task features, the main axes of variability extracted using these techniques often intermix task and nuisance variables, making them hard to interpret. Alternatively, dimensionality reduction techniques that do allow for estimating task-aligned axes of variability \cite{brendel2011demixed,semedo2019cortical,keeley2020,glaser2020recurrent}, do not apply to communication subspaces between brain areas, and/or necessitate trial repeat structure that does not occur in natural behavior.

Here, we introduce a probabilistic approach for learning interpretable task-relevant neural manifolds that capture both intra- and inter-area neural variability with single trial resolution. Task Aligned Manifold Estimation with Gaussian Process priors (TAME-GP) incorporates elements of demixed PCA (dPCA; \cite{machens2010demixing,brendel2011demixed,kobak2016demixed}) and probabilistic canonical correlation analysis (pCCA; \cite{bach2005probabilistic})\footnote{See Suppl.\ Info.\ S1 for background on probabilistic PCA, CCA and their relation to TAME-GP} into a graphical model that additionally includes biologically relevant Poisson noise. The model uses a Gaussian Process (GP) prior to enforce temporal smoothness, which allows for robust reconstruction of single-trial latent dynamics (see \cite{damianou2016multi} for a similar approach using Gaussian observation noise). We demonstrate the robustness and flexibility of TAME-GP in comparison to alternative approaches using synthetic data and neural recordings from monkeys performing a spatial navigation task in virtual reality. This reveals TAME-GP as a valuable tool for dissecting different sources of variability within and across brain areas during naturalistic behavior, with single-trial resolution.

\paragraph{Related work. }
Dimensionality reduction is usually achieved by unsupervised methods that identify axes of maximal variability in the data, such as PCA. In neuroscience, this is often accompanied by additional smoothing over time reflecting the underlying neural dynamics (e.g., Gaussian process factor analysis (GPFA) \cite{yu2008gaussian}; see GP-LVM \cite{ek2009shared} for similar approaches outside of neuroscience). This low dimensional projection is followed by a \textit{post hoc} interpretation of latents in the context of behavioral variables, often by visualization. Alternative approaches such as dPCA \cite{machens2010demixing,brendel2011demixed,kobak2016demixed} explicitly look for axes of neural variability that correlate with task variables of interest. However, these require partitioning trials into relatively few categories, based on experimental conditions or behavioral choices and averaging within conditions. This makes them unusable in naturalistic tasks where a single trial treatment is needed. Similarly, SNP-GPFA \cite{keeley2020} can partition (multi-region) neural activity into `shared signal' and `private noise' components, but only using data with stimulus repeats. Under `no-repeat' conditions, pCCA \cite{bach2005probabilistic} can find subspaces of maximal cross-correlation between linear projections of task variables and neural responses (under gaussian noise assumptions), without the need for \textit{a priori} grouping of trials by experimental condition or choice. This approach can also be applied for determining shared axes of co-variability across areas, an analog for communication subspaces \cite{semedo2019cortical}. Nonetheless, its noise model assumptions are mismatched to neural data. More fundamentally, pCCA only considers pairwise relationships, preventing a joint multi-area and task variables analysis. Overall, existing approaches come with practical limitations and do not directly address the routing of task-relevant information across brain areas.

\section{Task-aligned manifold estimation with GP priors (TAME-GP)}
\begin{figure}
\centering
\includegraphics[width=0.9\textwidth]{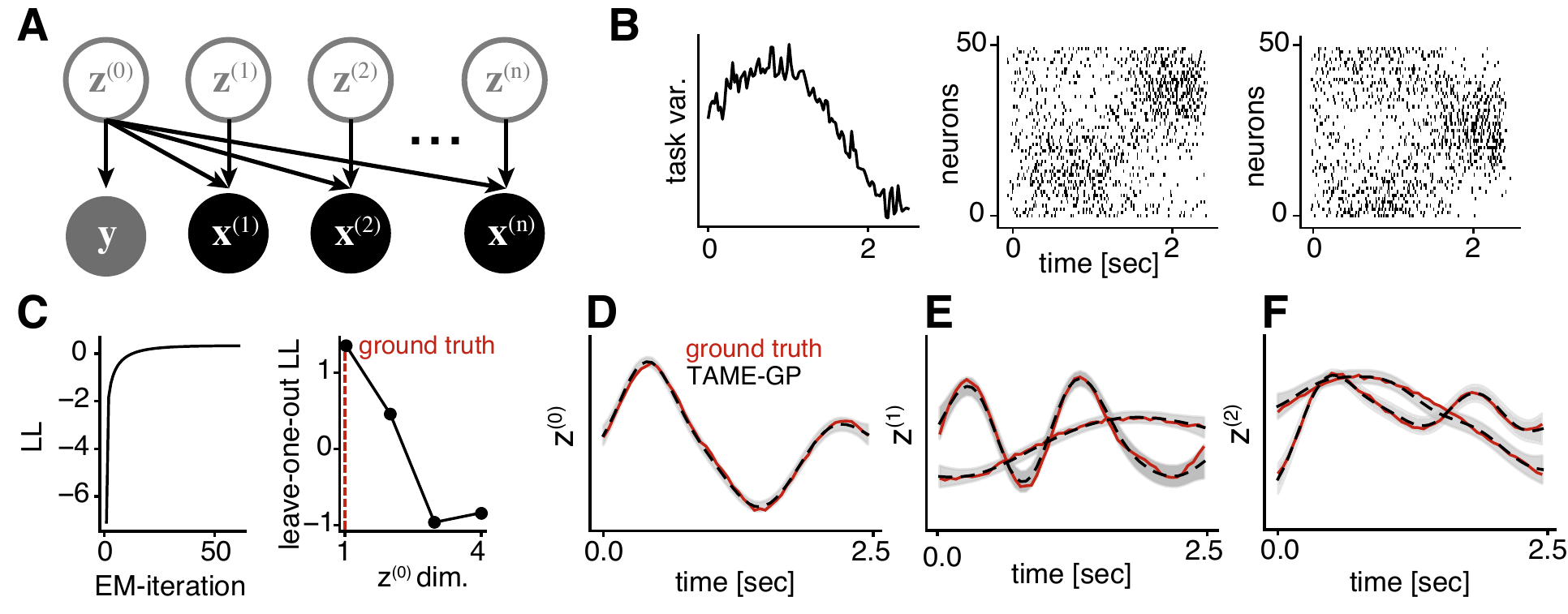}
\caption{
\textbf{A}. TAME-GP generative model. \textbf{B}. Example draws of spiking activity and a task variable from the TAME-GP graphical model.  
$\textbf{C}$. Model log-likelihood as a function of the EM iteration (left) and cross-validated leave-one-neuron-out marginal likelihood as a function of $\mathbf{z}^{(0)}$ dimension (right). \textbf{D}-\textbf{F}. Latent variables estimation for within model simulated data: ground truth latent factors and model posterior mean $\pm$ 95\% CI for three latent dimensions.}
\label{fig:model}
\end{figure}

In its most general form, the graphical model of TAME-GP models a set of spike-count population responses $\mathbf{x}^{(j)}$ from up to $n$ different areas,\footnote{Variables $\mathbf{x}^{(j)}$, $\mathbf{y}$ are tensors with dimensions corresponding to 1) an area-specific number of neurons/ task variable dimension, 2) time within trial, and 3) trial index.  We make indices explicit only where strictly needed.} together with task variable of interest $\mathbf{y}$ (Fig.~\ref{fig:model}A). The neural responses are driven by a set of $n+1$ low-dimensional latent variables $\mathbf{z^{(j)}}.$ Specifically, the responses in area $j$ arise as a linear combination of private latent variability $\mathbf{z}^{(j)}$ and shared latents $\mathbf{z}^{(0)}$, with Poisson noise and an exponential link function:
\begin{equation}
     p\left(\mathbf{x}^{(j)}_i | \mathbf{z}^{(0:n)}\right) = \text{Poisson}\left(\text{exp}\left(W^{(0,j)}_i \mathbf{z}^{(0)} + W^{(j,j)}_i \mathbf{z}^{(j)}+h_i^{(j)}\right)\right),
\end{equation}
with parameters $\mathbf{W}^{(0/j,j)}$ and $\mathbf{h}^{(j)}$.
     
To make the latents interpretable with respect to the task variables of interest, we adapt a probabilistic framing of CCA \cite{bach2005probabilistic} which introduces dependencies between one of the latents, for instance the shared component $\mathbf{z}^{(0)}$ and $\mathbf{y}$:
\begin{equation}
    p\left(\mathbf{y}|\mathbf{z}^{(0)}\right) = \mathcal{N}\left(\mathbf{y};\mathbf{C}\mathbf{z}^{(0)} + \mathbf{d},\boldsymbol{\Psi}\right),
\end{equation}
with parameters $\mathbf{C}$, $\mathbf{d}$ and $\boldsymbol{\Psi}$.

Finally, we regularize all latents to be smooth over time, through the introduction of a Gaussian Process prior, as in GPFA \cite{yu2008gaussian}:
\begin{equation}
    {z}^{(j)} \sim \text{GP}\left(\mathbf{0}, k_j(\cdot,\cdot)\right),
\end{equation}
with area and dimension specific hyperparameters $\tau$,
$
\text{k}\left(z^{(j)}_{t,i}, z^{(j)}_{t',i'}\right) =
    \delta_{ii'}\exp \left( -\frac{(t-t')^2}{2 \tau^{(j)}_i} \right),
$
where $z^{(j)}_{t,i}$ is the $i$-th component of the $j$-th latent at time $t$, and $\delta_{ii'}$ is the Kronecker delta.

Putting these elements together results in a factorization of the joint distribution of the form:
\begin{equation}
    p\left(\mathbf{x}^{(1:n)},\mathbf{y},\mathbf{z}^{(0:n)}\right) = 
    \prod_{j=0}^n p\left(\mathbf{z}^{(j)}\right) 
    p\left(\mathbf{y} | \mathbf{z}^{(0)}\right) 
    \prod_{i,j} p\left(x_i^{(j)}|\mathbf{z}^{(0)},\mathbf{z}^{(j)}\right). \label{eqn:jointLike}
\end{equation}

While this general form may not be completely intuitive at first pass, it allows for a unified mathematical treatment of several estimation tasks of interest. We will detail key instances of this class that have practical relevance for neuroscience when presenting our numerical results below.

\section{EM-based parameter learning}
\paragraph{E-step} Since a closed form solution of the posterior is not available (because of the Poisson noise model), we construct a Laplace approximation of the posterior \footnote{We group latents in $\mathbf{z}$, spike counts in $\mathbf{x}$ and $\boldsymbol{\theta} = \left\{\mathbf{W}^{(0/j,j)},\mathbf{h}^{(j)},\mathbf{C},\mathbf{d},\boldsymbol{\Psi},\mathbf{\tau}^{(j)}\right\}$, to  simplify notation.},  $p\left(\mathbf{z}|\mathbf{x},\mathbf{y},\boldsymbol{\theta}\right) \approx  q\left(\mathbf{z}|\mathbf{x},\mathbf{y},\boldsymbol{\theta}\right) = \mathcal{N}\left(\mathbf{z}; \mathbf{\hat{z}}, -\mathbf{H}^{-1}\right),
$
where $\hat{\mathbf{z}}$ is the MAP of the joint log-likelihood and $\mathbf{H}$ is its corresponding Hessian. Both of these quantities are estimated numerically.

The MAP estimate is obtained by gradient descent on the joint log likelihood. Using Eq.~(\ref{eqn:jointLike}), the gradient of the joint log likelihood w.r.t.\ the latents can be written as
\begin{alignat*}{2}
\nabla_{\mathbf{z}^{(j)}} \log p\left(\mathbf{z}, \mathbf{x}, \mathbf{y}\right) &= \sum_{l} \Big(\sum_{j\ge 0} \nabla_{\mathbf{z}^{(j)}} \log p\left(\mathbf{z}^{(j)}\right)
+\sum_{t > 0} \nabla_{\mathbf{z}^{(j)}} \log p\left(\mathbf{y}_t | \mathbf{z}_t^{(0)}\right)  \\
&+ \sum_{t > 0} \sum_{j>0} \nabla_{\mathbf{z}^{(j)}} \log p\left(\mathbf{x}^{(j)}_t | \mathbf{z}_t^{(0)}, \mathbf{z}_t^{(j)}\right)\Big),
\end{alignat*}
where $l\in (1:M)$ refers to the trial number, explicit index omitted for brevity. For a given trial, expanding one term at the time we have
\begin{align*}
    \nabla_{\mathbf{z}^{(j)}} \log p\left(\mathbf{z}^{(j)}\right) &= - \mathbf{K}^{(j)} \mathbf{z}^{(j)}\\
    \nabla_{\mathbf{z}^{(0)}_t} \log p\left(\mathbf{y} | \mathbf{z}^{(0)}_t \right) &=  \mathbf{C} \tr \Psi^{-1} \left( \mathbf{y}_t - \mathbf{C} \mathbf{z}^{(0)}_t -  \mathbf{d} \right)\\
    \nabla_{\mathbf{z}^{(k)}_t} \log p\left(\mathbf{x}^{(j)}_t | \mathbf{z}_t^{(0)}, \mathbf{z}_t^{(j)}\right) &=  \mathbf{W}^{(k,j)\top} \left( \mathbf{x}_t - \exp \left( \mathbf{W}^{(0,j)} \mathbf{z}^{(0)}_t +   \mathbf{W}^{(j,j)} \mathbf{z}^{(j)}_t + \mathbf{h}^{(j)}\right) \right),
\end{align*}
where $ \: j > 0,\: k\in \{0,j\}$.
The corresponding second moments are
\begin{alignat*}{2}
    \nabla^2_{\mathbf{z}^{(j)}} \log p\left(\mathbf{z}^{(j)}\right) &= - \mathbf{K}^{(j)} \; j \in (0:n)\\
    \nabla^2_{\mathbf{z}^{(0)}_t}\log p\left(\mathbf{y} | \mathbf{z}^{(0)}_t\right) &= -\mathbf{C}\tr \boldsymbol{\Psi}^{-1} \mathbf{C} \\
    \nabla_{\mathbf{z}_t^{(h)}}\nabla_{\mathbf{z}_t^{(k)}} \log p\left(\mathbf{x}^{(j)}_t | \mathbf{z}^{(0)}_t,\mathbf{z}^{(j)}_t\right) &= -\mathbf{W}^{(k,j) \top} \text{diag} \left( \exp \left( \mathbf{W}^{(0,j)} \mathbf{z}^{(0)}_t + \mathbf{W}^{(j,j)} \mathbf{z}^{(j)}_t + \mathbf{h}^{(j)}\right) \right)  \mathbf{W}^{(h,j)}.
\end{alignat*}
with $h,k \in \{0,j\}$.
Inverting the $D\times D$ dimensional Hessian matrix is cubic in $D = T \sum_j d_j$, where $T$ is the trial length and $d_j$ denotes the dimensionality of latent $\mathbf{z}^{(j)}$, which restricts the number and dimensionality of latents in practice.
The Hessian of the log likelihood is sparse but does not have a factorized structure. Nonetheless, we can take advantage of the block matrix inversion theorem, to speed up the computation to $\mathcal{O}(T^3\sum_j d_j^3 )$ (see Suppl.\ Info.\ S2 for  details), with additional improvements based on sparse GP methods \cite{wilson2015kernel,gardner2018gpytorch} left for future work.


\paragraph{M-step}
Given the approximate posterior $q$ found in the E-step,  the parameters updates can be derived analytically for a few parameters, and numerically for the rest. Introducing the notation $\bm{\mu}^{(k)}_t = \mathbb{E}_q[\bm{z}^{k}_t]$ and $\bm{\Sigma}^{(k,h)}_t = \mathbb{E}_q[\bm{z}^{(k)}_t \bm{z}^{(h) \top}_t] - \bm{\mu}^{(k)}_t \bm{\mu}^{(h)\top}_t$, we have
\begin{align*}
    \bar{\mathbf{C}} &= \left[ \sum_{l,t} \mathbf{y}_t\boldsymbol{\mu}^{(0)\top}_t -\frac{1}{T M}\sum_{l,t} \mathbf{y}_t \sum_{l,t} \boldsymbol{\mu}^{(0) \top}_t \right] \left[\sum_{l,t} \bm{\Sigma}^{(0,0)}_t + \sum_{l,t} \boldsymbol{\mu}^{(0)}_t \boldsymbol{\mu}^{(0)\top}_t -\frac{1}{T M}  \sum_{l,t} \boldsymbol{\mu}^{(0)}_t  \sum_{l,t} \boldsymbol{\mu}^{(0)\top}_t \right]^{-1}\\
    \bar{\mathbf{d}} &= \frac{1}{T M}\left( \sum_{l,t} \mathbf{y}_t -\bar{\mathbf{C}} \sum_{l,t} \boldsymbol{\mu}^{(0)}_t\right)\\
    \bar{\Psi} &= \frac{1}{T M}\left[ \sum_{l,t} \mathbf{y}_{t} \mathbf{y}_t\tr -\left(\sum_{l,t} \mathbf{y}_t \boldsymbol{\mu}^{(0)\top}_t \bar{\mathbf{C}}\tr+\bar{\mathbf{C}}\sum_{l,t}  \bm{\mu}^{(0)}_t\mathbf{y}_t\tr\right)  -\left( \sum_{l,t} \mathbf{y}_t \bar{\mathbf{d}}\tr+ \bar{\mathbf{d}}\sum_{l,t} \mathbf{y}_t\tr\right) \right. \nonumber\\
    &\left. +\bar{\mathbf{C}}\left( \sum_{l,t} (\bm{\Sigma}^{(0,0)}_t+ \boldsymbol{\mu}_t\bm{\mu}^{(0)}_t) \right)\bar{\mathbf{C}}\tr  + \left(\bar{\mathbf{C}}\sum_{l,t} \bm{\mu}^{(0)}_t  \bar{\mathbf{d}}\tr + \bar{\mathbf{d}} \sum_{l,t} \bm{\mu}^{(0)\top}_t \bar{\mathbf{C}}\tr \right)+TM \bar{\mathbf{d}}\bar{\mathbf{d}}\tr \right]
\end{align*}
where $l=1:M$ and $t=1:T$ are trial and time within trial indices.

The other observation model parameters are computed numerically by optimizing the expected log-likelihood under the posterior. In particular, for neuron $i$ in population $j$ we have
\begin{align}
    \mathcal{L} \left( W^{(0,j)}_i , W^{(j,j)}_i,h_i \right) &= \sum_{t,l} x_{ti} \left(h_i + \begin{bmatrix}
    W^{(0,j)}_i & W^{(j,j)}_i 
    \end{bmatrix} 
    \begin{bmatrix}
    \bm{\mu}^{(0)}_t \\
    \bm{\mu}^{(j)}_t
    \end{bmatrix} \right)\nonumber\\
    &- \exp \left( h_i + 
    \begin{bmatrix}
    W^{(0,j)}_i & W^{(j,j)}_i 
    \end{bmatrix} 
    \begin{bmatrix}
    \bm{\mu}^{(0)}_t  \\
    \bm{\mu}^{(j)}_t
    \end{bmatrix} \right.\nonumber\\
    &+ \left. \frac{1}{2} 
    \begin{bmatrix}
     W^{(0,j)}_i & W^{(j,j)}_i 
    \end{bmatrix}
    \begin{bmatrix}
    \bm{\Sigma}^{(0,0)}_t & \bm{\Sigma}^{(0,j)}_t \\
    \bm{\Sigma}^{(0,j)\top}_t & \bm{\Sigma}^{(j,j)}_t 
    \end{bmatrix}
    \begin{bmatrix}
    W^{(0,j) \top}_i \\ 
    W^{(j,j) \top}_i 
    \end{bmatrix} \right).
\end{align}
 For each neural population, we jointly optimized the projection weights and the intercept of all neurons with a full Newton scheme by storing the inverse Hessian in compressed sparse row (CSR) format (see Suppl.Info. S3 for the gradient and Hessian of $\mathcal{L}$).
 
The GP-prior parameters were also learned from data by gradient based optimization (using the limited-memory Broyden–Fletcher–Goldfarb–Shanno scheme \cite{2020SciPy-NMeth}). First, we set $\lambda^{(j)}_i  = -\log (2 \tau^{(j)}_i)$, and optimize for $\lambda^{(j)}_i$ to enforce a positive time constant. We define $\bm{K}^{(j)}_i\in \mathbb{R}^{T \times T}$, such that $\left[\mathbf{K}^{(j)}_{i}\right]_{ts} = \exp \left( - e^{\lambda^{(j)}_i} (t-s)^2 \right)$. The resulting objective function will take the form,
$
    \mathcal{L}\left(\lambda^{(j)}_i \right) = -\text{trace} \left(\bm{K}^{(j) -1}_i \mathbb{E}_q[\bm{z}^{(j)}_i \bm{z}^{(j)\top}_i] \right) - \log |\bm{K}^{(j)}_i|.
$
Gradients are provided in Suppl.~Info S4.

\paragraph{Parameter initialization.}
Since EM is only guaranteed to converge to a local optimum, the quality of the final estimate depends significantly on the choice of initialization. To address this, we use estimates from a factorized version of TAME-GP, where the GP prior is replaced with a factorized normal distribution. This  temporal independence assumption allows for an efficient inversion of the model posterior covariance that remains sparse and can be stored as a compressed sparse row matrix. This allows to efficiently optimize the Poisson observation parameters via a full Newton scheme. The EM estimates of the factorized TAME are themselves initialized using canonical correlation analysis (CCA), which we found to improve the initial marginal likelihood over alternative initializations, despite mismatched model assumptions. See Suppl.~Info S5. for detail.  

\section{Results}
\paragraph{Latent reconstruction for within model data.}
To validate the estimation procedure, we first used a simulated dataset sampled from the TAME-GP graphical model, with predefined parameters. Specifically, we simulated two neural populations $\mathbf{x}^{(1)}$ and $\mathbf{x}^{(2)}$, each with 50 units and a one-dimensional task relevant variable $y$. We fixed the private latent factors $\mathbf{z}^{(1)}$ and $\mathbf{z}^{(2)}$ to two dimensions, and that of the shared factor $\mathbf{z}^{(0)}$ to one. The projection weights $\mathbf{W}^{(j)}$ and $\mathbf{C}$, the intercept terms $\mathbf{d}$ and $\mathbf{h}^{(j)}$, the observation variance matrix $\Phi$, and the GP time constants of the factors were randomly assigned. The parameters were chosen such that the overall mean firing rate was about 20Hz in both areas. We simulated spike counts at 50ms resolution for 200 trials, each lasting  2.5 seconds (see example trial in Fig.~\ref{fig:model}B). Given this data, we assessed the ability of our EM-based estimator to recover its true latent structure.\footnote{Here and in all subsequent analyses 90\% of the data is used for training the model and 10\% for testing.}
The marginal log likelihood saturated after a relatively small number of EM iterations (Fig.~\ref{fig:model}C). As a basic test of our ability to determine the dimensionality of latents, we systematically varied the dimensionality of the shared latent, while fixing the dimensions of $\mathbf{z}^{(1)}$ and $\mathbf{z}^{(2)}$ to their ground truth value of 2. We found that the best model fit was achieved at the ground truth task dimension 1, demonstrating that we are able to infer true latent dimensionality from data (Fig.\ref{fig:model}D).

Finally, we assessed the quality of the recovered latents in individual test trials. Due to known degeneracies, originally documented in linear gaussian latent models \cite{roweis1999unifying}, the latent factors in TAME-GP are identifiable up to an affine transformation of the latent space. To address this, we used Procustes \cite{schonemann1966generalized} to realign the latent axes back to the original space. The resulting posterior mean estimate of the latents show an excellent agreement with the ground truth factors (cross-validated linear regression $R^2$ of 0.99 between the MAP estimate of latents and ground truth, Fig.~\ref{fig:model} D-F), while the model predicted rates explained 98\% of the ground truth firing rate variance. Overall, these numerical tests confirm that EM provides a veridical estimation of ground truth latent structure for within distribution data.

\paragraph{Task-aligned latent reconstruction for simulated latent dynamical systems models.}
The simple graphical model of TAME-GP captures axes of neural variability of scientific interest, but is far from an accurate generative model for neural dynamics during behavior. To assess the ability of TAME-GP to extract underlying structure from complex and out-of-distribution neural data, we used latent dynamical systems models in which we can explicitly define the flow of information from external stimuli and between areas, in several scenarios of practical interest.

\begin{figure}
\centering
\includegraphics[width=0.95\textwidth]{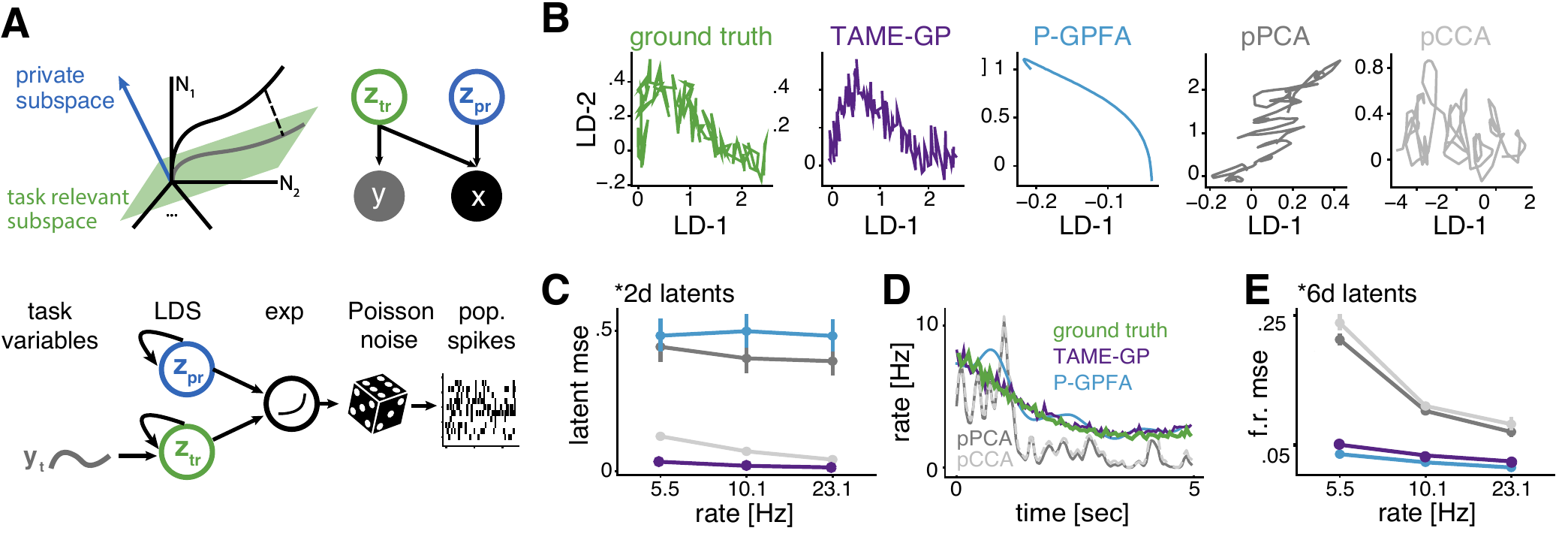}
\caption{Methods comparison for single area task manifold alignment. $\textbf{A}$. TAME-GP graphical model for single area (top) and schematic for data generating process (bottom). \textbf{B}. Ground truth task relevant dynamics (green) and estimated low dimensional  projection for TAME-GP (purple), P-GPFA (blue), pPCA (dark gray) and pCCA (light gray).\textbf{C} Mean squared error between the true shared dynamics and the model reconstruction, mean $\pm$ s.d. over 10-fold cross-validation. \textbf{D}. Example single trial firing rate reconstruction. \textbf{E}. Mean squared error between the true and reconstructed firing rate across conditions, mean $\pm$ s.d. over 10-folds of cross-validation.}
\label{fig:task_rel_dyn}
\end{figure}

The first \textit{in silico} experiment focuses on identifying axes of task-relevant variability in neural responses. As a simple test case, we modeled a single neural population with a 6d latent structure (Fig.~\ref{fig:task_rel_dyn}A). Two of the latent dimensions were task-relevant, driven by an observed temporally smooth external input $\mathbf{y}_t$, while the other four dimensions were completely intrinsic to the circuit. The key distinction between this process and the TAME-GP model assumptions is that the observed task variable acts as an input drive to the underlying latent dynamics rather than mapping to the latents directly. The latent dynamics take the form of a multivariate AR(1),
\begin{align}
\begin{cases}
  \mathbf{z}_{\text{pr},t+1} &= A_{\text{pr}} \left(\mathbf{z}_{\text{pr},t} - \mathbf{\mu}_t \right) \Delta t + \sqrt{2\Delta t} \;\text{d} \mathbf{w}^{(0)}_{t}  \\
    \mathbf{z}_{\text{tr},t+1} &= A_{\text{tr}} \left(\mathbf{z}_{\text{tr},t} - \mathbf{y}_t \right) \Delta t + \sqrt{2\Delta t} \;\text{d} \mathbf{w}^{(1)}_{t},
    \end{cases}
\end{align}
where $A_{\text{pr}} \in \mathbb{R}^{4 \times 4}$ and $A_{\text{tr}} \in \mathbb{R}^{2 \times 2}$ the private and task relevant dynamics, $\mathbf{y}_t \in \mathbb{R}^2$ and $ \mathbf{\mu}_t \in\mathbb{R}^4$ inputs drawn from a factorized RBF kernel, and $\bm{w}^{(i)}_{t}$ is independent white noise for $i=0,1$. Given these latent dynamics, spikes are generated as described by the TAME-GP observation model with $\mathbf{W}\in\mathbb{R}^{100 \times 6}$, and $\mathbf{d} \in \mathbb{R}^{100}$. We adjusted the parameters as to cover several average population firing rates by regulating $\mathbf{d}$, for a fixed number of trials (200) and a fixed trial duration (5 seconds). For simplicity, we circumvent the hyperparameter selection step by assuming that all estimators have access to the ground truth latent dimensionality: TAME-GP assumed 2 shared and 4 private latents. Unsupervised methods (pPCA, P-GPFA) were tasked with extracting the main two axes of neural variability in the data, while the supervised methods (pCCA) estimated 2d latents that correlate with task variable $\mathbf{y}$; the same alignment procedure was used to align the resulting axes to the original in all cases.

Fig.~\ref{fig:task_rel_dyn}B illustrates the latent dynamics as estimated by TAME-GP, pPCA \cite{tipping1999probabilistic}, P-GPFA \cite{hooram2015poisson}, and pCCA \cite{bach2005probabilistic} . We quantify the latent space estimation accuracy by mean squared error, demonstrating that TAME-GP captured the stimulus driven dynamics better than other methods (Fig.~\ref{fig:task_rel_dyn}C, see Suppl.~Info. Fig. S1). P-GPFA showed a tendency to over-smooth, which obscured most of the underlying fine timescale latent structure. PCA failed by focusing on main axes of variability irrespective of task relevance, while CCA estimates were visually less interpretable. Only pCCA and TAME-GP found projections that selectively encoded for $\mathbf{z}_{tr}$ with TAME-GP outperforming pCCA across conditions.

We also compared these methods in terms of their ability to predict the ground truth firing rate generating the observed spiking responses (total dimensions matching the ground truth of 6). Both TAME-GP and P-GPFA showed a stable and accurate firing rate reconstruction error across conditions (Fig.~\ref{fig:task_rel_dyn}D,E), while the factorized linear gaussian methods (pPCA, pCCA) performed poorly. This may be due to the larger model mismatch, while additionally suffering from the lack of temporal smoothing, especially for low firing rates. Overall, TAME-GP was the only procedure that both captured the overall data statistics well and extracted accurate task-interpretable latents.
 
\begin{figure}[t]
\centering
\includegraphics[width=0.85\textwidth]{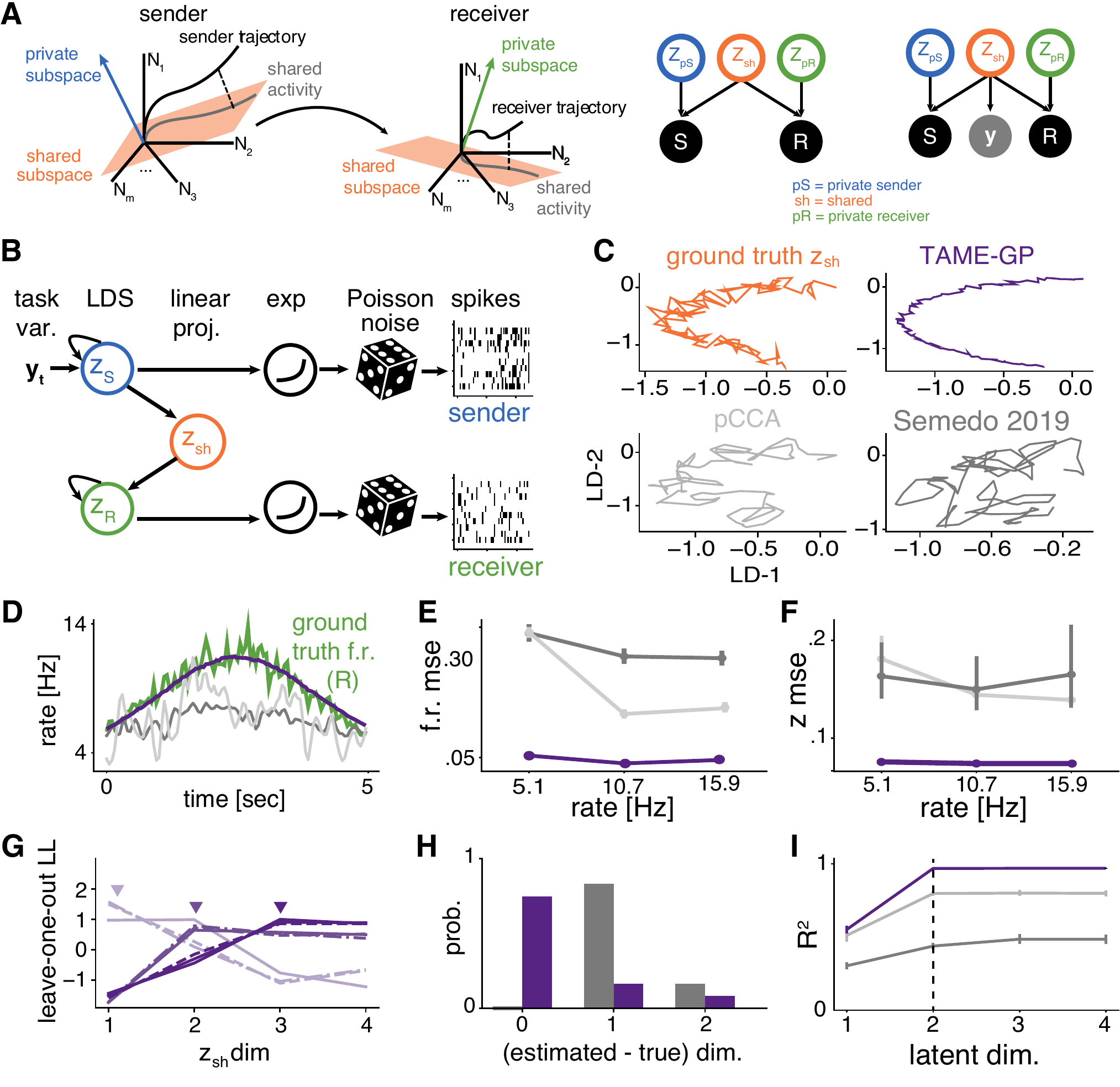}
\caption{$\textbf{A}$. Schematic of communication subspace (left) and associated TAME-GP graphical model versions (right).
\textbf{B}. Ground truth spike count generation process. 
\textbf{C}. Example shared latent reconstruction for TAME-GP (purple), PCCA (light grey) and, reduced rank regression (dark grey); ground truth in orange. 
\textbf{D}. Example reconstructions of the receiver firing rates compared to the ground truth (green) and \textbf{E}. corresponding summary statistics for different mean firing rates.
\textbf{F}. Statistics of shared dynamics reconstruction. 
\textbf{G}. TAME-GP leave-one-neuron-out log-likelihood for different ground truth shared manifold dimensionality (d=1,2,3) and increasing population rate from 5.1, 10.7, 15.9 Hz (respectively, dashed, dashed-dotted and continuous lines). Lines styles show different average firing rate conditions. 
\textbf{H}. Difference between estimated and true $\mathbf{z}_{\text{sh}}$ dimensionality for TAME-GP (purple) and reduced rank regression (grey). 
\textbf{I}. Model fit quality as a function of the number of dimensions for the different estimators. Ground truth dimension d=2 (dashed line). Error bars show mean $\pm$ s.d. over 10-folds of cross-validation.}
\label{fig:comm}
\end{figure}

\paragraph{Assessing inter-area communication in simulated latent dynamical systems} In the second set of numerical experiments, we focused on estimating low-dimensional communication sub-spaces across neural populations (Fig.~\ref{fig:comm}A). The ground truth data was again constructed using latent dynamical systems models, which now included two populations (Fig.~\ref{fig:comm}B), where a low dimensional projection of the dynamics in one area, the sender, drive the dynamics of the other area, the receiver:
\begin{align}
\begin{cases}
\mathbf{z}_{\text{S},t+1} &= A_{S} \left(\mathbf{z}_{\text{S},t} - \mathbf{y}_t \right) \Delta t + \sqrt{2\Delta t} \mathbf{w}^{(0)}_t  \\
\mathbf{z}_{\text{sh}} &= P \cdot \mathbf{z}_{\text{S}} \\
\mathbf{z}_{\text{R},t+1} &= A_{R} \left(\mathbf{z}_{\text{R},t} - \mathbf{\lambda}_t - \mathbf{z}_{\text{sh},t}\right) \Delta t + \sqrt{2\Delta t}  \mathbf{w}^{(1)}_t ,
\end{cases}
\end{align}
where $A_S \in \mathbb{R}^{4 \times 4}$ and $A_R\in \mathbb{R}^{4\times 4}$ are the sender and receiver dynamics, $\mathbf{y}_t$ and $\mathbf{\lambda}_t$ are temporally smooth inputs drawn from independent GPs with factorized RBF kernels, $P \in \mathbb{R}^{2\times 4}$ defines the shared submanifold projection, and $w^{(i)}_{t}$ is independent white noise. These latents map into spikes as above. We simulated three average firing rate conditions and varied the ground truth number of shared dimensions, from one to three. We compared our method with two commonly used alternatives: pCCA and Semedo's reduced-rank regression procedure for communication manifold estimation \cite{semedo2019cortical} (Fig.~\ref{fig:comm}C), as well as with SNP-GPFA \cite{keeley2020} (both with and without trial repeats, see Suppl.Info. Section S6 and Fig. S2). 

TAME-GP (without task alignment) outperformed alternative approaches in terms of the reconstruction error of both ground truth firing rates (Fig.~\ref{fig:comm}D,E) and shared latent dynamics (Fig.~\ref{fig:comm}F). Furthermore, when testing the ability of different approaches to infer the dimensionality of the shared manifold through model comparison, the leave-one-out likelihood saturated at the ground truth dimension for all simulations (Fig.~\ref{fig:comm}G), and peaked at the correct dimension 75\% of the times (Fig.~\ref{fig:comm}H). In contrast, the Semedo estimator tended to systematically overestimate the dimensionality of the shared manifold. Finally, we also tested the general case in which we search for a communication subspace that aligns to task variable $\mathbf{y}$. To do so, we fit TAME-GP to the same dataset but assuming that $\mathbf{y}_t$ is observed. We found again that TAME-GP has the best reconstruction accuracy, which saturates at the ground truth dimensionality (d=2). These observations are consistent across firing rate levels (see Suppl.~Info. Fig. S3). For the SNP-GPFA comparison, we find that in the case of precise stimulus repetitions both models are able to capture the latent space factorization. Instead, only TAME-GP generalizes to the case where latent dynamics vary trial to trial (see Suppl.~Info. Fig.~S2, Table S1 and Section S6 for details). Overall, these results suggest that TAME-GP can robustly recover meaningful sources of co-variability across areas.

\begin{figure}
\centering
\includegraphics[width=0.9\textwidth]{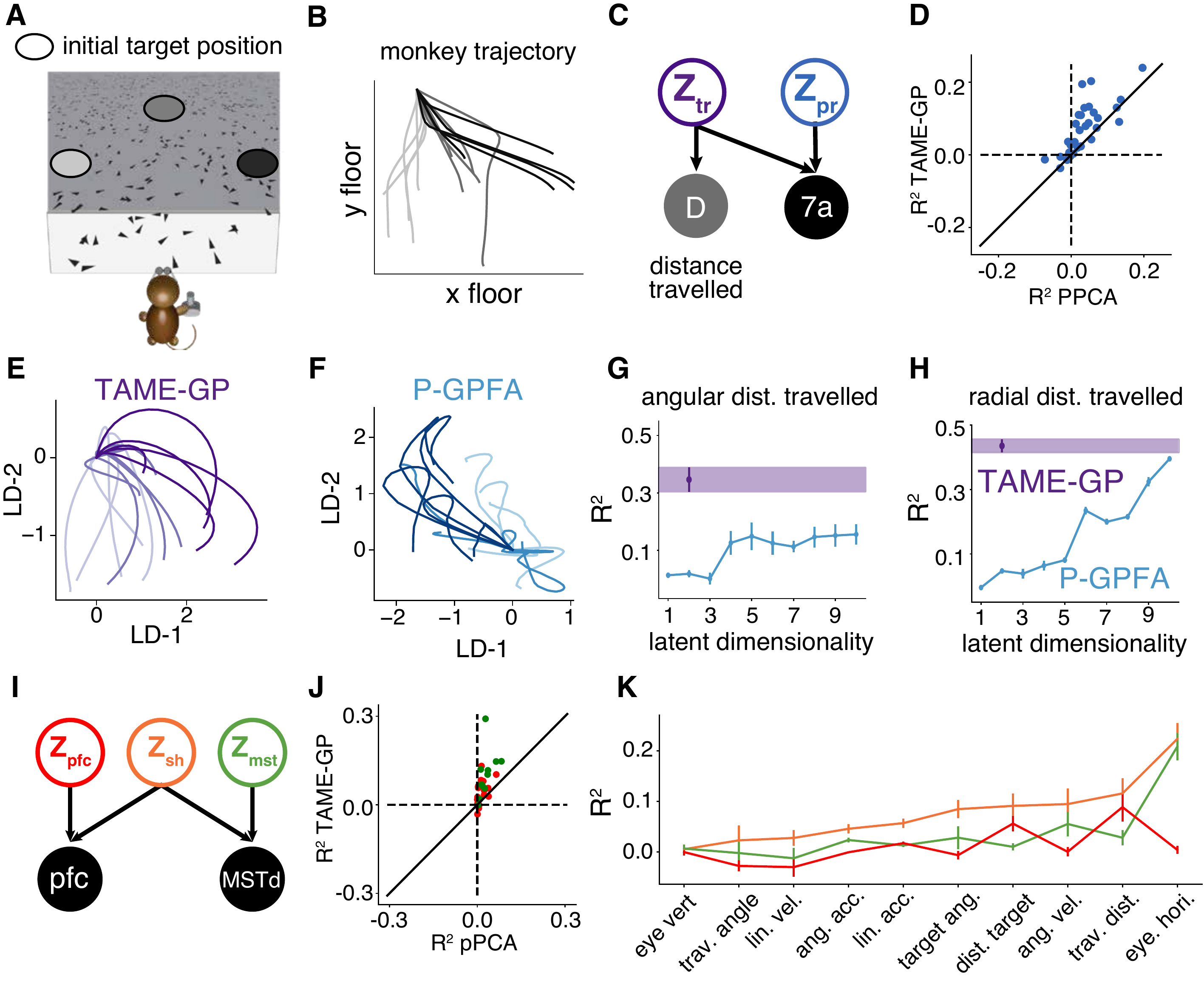}
\caption{
Analysis of macaque neural recordings. 
$\textbf{A}$. Schematic of task. Initial target location is randomized and remains visible for 300ms. The monkey has to use the joystick to navigate to the believed target position.
$\textbf{B}$. Top view of example monkey trajectories; increasing contrast marks initial location of the target (right, center, left). $\textbf{C}$. Within-area TAME-GP estimation aligned a latent task variable: the distance travelled.\textbf{D}. Scatter plot of leave-one-neuron-out spike count variance explained for dimension-matched TAME-GP and pPCA. Dots represent individual neurons. 
\textbf{E}. Single trial TAME-GP estimates of the task relevant dynamics, compared to \textbf{F}. those of P-GPFA. Trajectories are color-graded according to the initial angular target location (as in B). Ridge regression decoding of \textbf{G}. and \textbf{H}. linear distance travelled. TAME-GP decoding $R^2$ (purple) is based on a two dimensional task relevant latent. P-GPFA $R^2$ (light blue) estimates were obtained for a range of latent dimensions (1-10). \textbf{I}. Communication subspace estimation between MSTd and dlPFC. 
\textbf{J}. As D, for shared latent space. \textbf{K}. Ridge regression decoding of task relevant variables (sorted by their shared subspace information content) from the shared (orange) and private latents (green, red) estimated by TAME-GP. Mean $R^2$ $\pm$ s.e.m. were computed with by 10-fold cross-validation.}
\label{fig:monkey_data}
\end{figure}

\paragraph{Multi-area neural recordings in monkeys}
We tested the ability of TAME-GP to find task aligned neural manifolds in an experimental dataset characterized by a high-dimensional input space and the lack of trial repeats. Specifically, macaques navigate in virtual reality by using a joystick controlling their linear and angular velocity to ``catch fireflies''(Fig.\ref{fig:monkey_data}A, B) \cite{lakshminarasimhan2018dynamic}. Spiking activity is measured (binned in 6ms windows, sessions lasting over 90min) and neurons in the two recorded brain areas (MSTd and dlPFC) showed mixed selectivity, encoding a multitude of task relevant variables \cite{noel2021flexible}. As a result, responses are high dimensional and unsupervised dimensionality reduction methods inevitably capture an uninterpretable mixture of task relevant signals in their first few latent dimensions. 

As a first test of the estimator, we used TAME-GP to extract latent projection that align with the ongoing distance from the origin, decomposed in an angular and a radial component (Fig.~\ref{fig:monkey_data}C). We set the task relevant latent $\mathbf{z}^{(0)}$ dimensions to two, matching the number of task variables. 
We verified the accuracy of the model by computing leave-one-neuron-out firing rate predictions and calculating the $R^2$ between model predictions and raw spike counts (as in \cite{yu2008gaussian}). This TAME-GP estimator systematically outperformed pPCA with matched number of latents by this metric (Fig.~\ref{fig:monkey_data}D). We also compared the latent factors found by TAME-GP to those obtained by P-GPFA (Fig.~\ref{fig:monkey_data}E,F). We asked (in $R^2$ terms) how much information about the task variables can be linearly decoded from their respective latents (Fig.~\ref{fig:monkey_data}G,H). For both variables, we found that the target variables were better accounted for by a two-dimensional TAME-GP estimated latent than by up to 10 dimensional latent spaces extracted with P-GPFA. This result shows that TAME-GP finds compact low dimensional accounts of neural variability with respect of task variables of interest.

Lastly, we probed the model's ability to learn a communication subspace (Fig.~\ref{fig:monkey_data}I) between MSTd and dlPFC, brain areas that are known to interact during this task \cite{noel2021flexible}). In this instance, we selected the number of shared and private latent dimensions by maximizing the leave-one-neuron-out spike counts variance explained over a  grid of candidate values (see Suppl.~Info. Fig. S4 and Section S7).  As before, we find that the TAME-GP reconstruction accuracy surpasses that of dimensionality-matched pPCA, for both MSTd and dlPFC (Fig.~\ref{fig:monkey_data}J). Since the shared manifold estimation was agnostic to task variables in this case, we used decoding from latent spaces to ask if the shared variability between these areas carried information about task variables known to drive single neuron responses in these areas. We found that the monkey's horizontal eye position, as well as latent task variables such as the travelled distance or the distance still remaining to target were mostly accounted for in shared, as opposed to private, axes of variability (Fig.~\ref{fig:monkey_data}K). This recapitulates prior observations made at the single-cell level (\cite{noel2021flexible}). Overall, the results demonstrate that TAME-GP can extract interpretable low-dimensional latents and shared neural subspaces from complex and high-dimensional datasets.

\section{Discussion}
Technological advances in systems neuroscience place an ever-increasing premium on the ability to concisely describe high-dimensional task-relevant neural responses. Here we introduce TAME-GP, a flexible statistical framework for partitioning neural variability in terms of private or shared (i.e., inter-area) sources, aligned to task variables of interest, and with single trial resolution. Our method was shown to provide compact latent manifold descriptions that better capture neural variability than any of the standard approaches we compared it against.

An important nuance that distinguishes various neural dimensionality reduction methods is whether the covariability being modeled is that of trial-averaged responses (i.e. stimulus correlations), residual fluctuations around mean responses (i.e. noise correlations) or a combination of the two (total correlations). Since isolating either the signal or the noise correlations alone would require across trial averages, our approach models total correlations, time resolved within individual trials. This differentiates our shared variability estimates from the traditional definition of a communication subspace \cite{semedo2019cortical}, which uses noise correlations alone, while keeping some of its spirit. It also makes it applicable to datasets without trial repeats. 

The model adapts the approach of pCCA as a way of ensuring that the extracted latents reflect axes of neural variability that carry specific task relevant information. This choice has appealing mathematical properties in terms of unifying the problems of finding interpretable axes and communication subspaces, but is not the most natural one in terms of the true generative process of the data. While behavioral outputs are causal outcomes of the neural activity as described by the TAME-GP graphical model, sensory variables act as drivers for the neural responses and should causally affect the latent dynamics, not the other way around. Hence a natural next step will be to incorporate in the framework explicit stimulus responses, perhaps by taking advantage of recent advances in estimating complex tuning functions during naturalistic behavior  \cite{balzani2020efficient}.

Similarly, the choice of temporal smoothing by RBF kernel GP was made for simplicity. It would be interesting to explore the use of priors with more interesting structure, for instance spectral mixture kernels \cite{wilson2013gaussian}, introducing prior dependencies across latent dimensions \cite{de2021mogptk}, or using non-reversible GP priors that better capture the causal structure of neural dynamics \cite{rutten2020non}. More generally, the probabilistic formulation allows the ideas formalized by TAME-GP to be combined with other probabilistic approaches for describing stimulus tuning and explicit latent neural dynamics \cite{zhao2017variational,nassar2018learning,duncker2019learning,glaser2020recurrent,duncker2021dynamics}. Hence, this work adds yet another building block in our statistical arsenal for tackling questions about neural population activity as substrate for brain computation.

\paragraph{Broader impact} 
We do not foresee any negative consequences to society from our work. Code for TAME-GP is available at: \emph{https://github.com/BalzaniEdoardo/TAME-GP} 



\small
\bibliography{neurips2022_refs}

\begin{thebibliography}{10}

\bibitem{michaiel2020dynamics}
Angie~M Michaiel, Elliott~TT Abe, and Cristopher~M Niell.
\newblock Dynamics of gaze control during prey capture in freely moving mice.
\newblock {\em Elife}, 9:e57458, 2020.

\bibitem{shamash2021mice}
Philip Shamash, Sarah~F Olesen, Panagiota Iordanidou, Dario Campagner, Nabhojit
  Banerjee, and Tiago Branco.
\newblock Mice learn multi-step routes by memorizing subgoal locations.
\newblock {\em Nature Neuroscience}, 24(9):1270--1279, 2021.

\bibitem{noel2021flexible}
Jean-Paul Noel, Edoardo Balzani, Eric Avila, Kaushik Lakshminarasimhan,
  Stefania Bruni, Panos Alefantis, Cristina Savin, and Dora~E Angelaki.
\newblock Flexible neural coding in sensory, parietal, and frontal cortices
  during goal-directed virtual navigation.
\newblock {\em bioRxiv}, 2021.

\bibitem{stevenson2011advances}
Ian~H Stevenson and Konrad~P Kording.
\newblock How advances in neural recording affect data analysis.
\newblock {\em Nature neuroscience}, 14(2):139--142, 2011.

\bibitem{jun2017fully}
James~J Jun, Nicholas~A Steinmetz, Joshua~H Siegle, Daniel~J Denman, Marius
  Bauza, Brian Barbarits, Albert~K Lee, Costas~A Anastassiou, Alexandru Andrei,
  {\c{C}}a{\u{g}}atay Ayd{\i}n, et~al.
\newblock Fully integrated silicon probes for high-density recording of neural
  activity.
\newblock {\em Nature}, 551(7679):232--236, 2017.

\bibitem{angotzi2019sinaps}
Gian~Nicola Angotzi, Fabio Boi, Aziliz Lecomte, Ermanno Miele, Mario Malerba,
  Stefano Zucca, Antonino Casile, and Luca Berdondini.
\newblock Sinaps: An implantable active pixel sensor cmos-probe for
  simultaneous large-scale neural recordings.
\newblock {\em Biosensors and Bioelectronics}, 126:355--364, 2019.

\bibitem{boi2020multi}
Fabio Boi, Nikolas Perentos, Aziliz Lecomte, Gerrit Schwesig, Stefano Zordan,
  Anton Sirota, Luca Berdondini, and Gian~Nicola Angotzi.
\newblock Multi-shanks sinaps active pixel sensor cmos probe: 1024
  simultaneously recording channels for high-density intracortical brain
  mapping.
\newblock {\em bioRxiv}, page 749911, 2020.

\bibitem{steinmetz2021neuropixels}
Nicholas~A Steinmetz, Cagatay Aydin, Anna Lebedeva, Michael Okun, Marius
  Pachitariu, Marius Bauza, Maxime Beau, Jai Bhagat, Claudia B{\"o}hm, Martijn
  Broux, et~al.
\newblock Neuropixels 2.0: A miniaturized high-density probe for stable,
  long-term brain recordings.
\newblock {\em Science}, 372(6539):eabf4588, 2021.

\bibitem{cunningham2014dimensionality}
John~P Cunningham and M~Yu Byron.
\newblock Dimensionality reduction for large-scale neural recordings.
\newblock {\em Nature neuroscience}, 17(11):1500--1509, 2014.

\bibitem{brendel2011demixed}
Wieland Brendel, Ranulfo Romo, and Christian~K Machens.
\newblock Demixed principal component analysis.
\newblock {\em Advances in neural information processing systems}, 24, 2011.

\bibitem{semedo2019cortical}
Jo{\~a}o~D Semedo, Amin Zandvakili, Christian~K Machens, M~Yu Byron, and Adam
  Kohn.
\newblock Cortical areas interact through a communication subspace.
\newblock {\em Neuron}, 102(1):249--259, 2019.

\bibitem{keeley2020}
Keeley S.L. Aoi M.C. Yu Y. Smith S.L.~Pillow J.W.
\newblock Identifying signal and noise structure in neural population activity
  with gaussian process factor models.
\newblock {\em NeurIPS}, 34, 2020.

\bibitem{glaser2020recurrent}
Joshua Glaser, Matthew Whiteway, John~P Cunningham, Liam Paninski, and Scott
  Linderman.
\newblock Recurrent switching dynamical systems models for multiple interacting
  neural populations.
\newblock {\em Advances in neural information processing systems},
  33:14867--14878, 2020.

\bibitem{machens2010demixing}
Christian~K Machens.
\newblock Demixing population activity in higher cortical areas.
\newblock {\em Frontiers in computational neuroscience}, 4:126, 2010.

\bibitem{kobak2016demixed}
Dmitry Kobak, Wieland Brendel, Christos Constantinidis, Claudia~E Feierstein,
  Adam Kepecs, Zachary~F Mainen, Xue-Lian Qi, Ranulfo Romo, Naoshige Uchida,
  and Christian~K Machens.
\newblock Demixed principal component analysis of neural population data.
\newblock {\em Elife}, 5:e10989, 2016.

\bibitem{bach2005probabilistic}
Francis~R Bach and Michael~I Jordan.
\newblock A probabilistic interpretation of canonical correlation analysis.
\newblock Technical report, 2005.

\bibitem{damianou2016multi}
Andreas Damianou, Neil~D Lawrence, and Carl~Henrik Ek.
\newblock Multi-view learning as a nonparametric nonlinear inter-battery factor
  analysis.
\newblock {\em arXiv preprint arXiv:1604.04939}, 2016.

\bibitem{yu2008gaussian}
Byron~M Yu, John~P Cunningham, Gopal Santhanam, Stephen Ryu, Krishna~V Shenoy,
  and Maneesh Sahani.
\newblock Gaussian-process factor analysis for low-dimensional single-trial
  analysis of neural population activity.
\newblock {\em Advances in neural information processing systems}, 21, 2008.

\bibitem{ek2009shared}
Carl~Henrik Ek and PHTND Lawrence.
\newblock {\em Shared Gaussian process latent variable models}.
\newblock PhD thesis, Citeseer, 2009.

\bibitem{wilson2015kernel}
Andrew Wilson and Hannes Nickisch.
\newblock Kernel interpolation for scalable structured gaussian processes
  (kiss-gp).
\newblock In {\em International conference on machine learning}, pages
  1775--1784. PMLR, 2015.

\bibitem{gardner2018gpytorch}
Jacob Gardner, Geoff Pleiss, Kilian~Q Weinberger, David Bindel, and Andrew~G
  Wilson.
\newblock Gpytorch: Blackbox matrix-matrix gaussian process inference with gpu
  acceleration.
\newblock {\em Advances in neural information processing systems}, 31, 2018.

\bibitem{2020SciPy-NMeth}
Pauli Virtanen, Ralf Gommers, Travis~E. Oliphant, Matt Haberland, Tyler Reddy,
  David Cournapeau, Evgeni Burovski, Pearu Peterson, Warren Weckesser, Jonathan
  Bright, St{\'e}fan~J. {van der Walt}, Matthew Brett, Joshua Wilson, K.~Jarrod
  Millman, Nikolay Mayorov, Andrew R.~J. Nelson, Eric Jones, Robert Kern, Eric
  Larson, C~J Carey, {\.I}lhan Polat, Yu~Feng, Eric~W. Moore, Jake
  {VanderPlas}, Denis Laxalde, Josef Perktold, Robert Cimrman, Ian Henriksen,
  E.~A. Quintero, Charles~R. Harris, Anne~M. Archibald, Ant{\^o}nio~H. Ribeiro,
  Fabian Pedregosa, Paul {van Mulbregt}, and {SciPy 1.0 Contributors}.
\newblock {{SciPy} 1.0: Fundamental Algorithms for Scientific Computing in
  Python}.
\newblock {\em Nature Methods}, 17:261--272, 2020.

\bibitem{roweis1999unifying}
Sam Roweis and Zoubin Ghahramani.
\newblock A unifying review of linear gaussian models.
\newblock {\em Neural computation}, 11(2):305--345, 1999.

\bibitem{schonemann1966generalized}
Peter~H Sch{\"o}nemann.
\newblock A generalized solution of the orthogonal procrustes problem.
\newblock {\em Psychometrika}, 31(1):1--10, 1966.

\bibitem{tipping1999probabilistic}
Michael~E Tipping and Christopher~M Bishop.
\newblock Probabilistic principal component analysis.
\newblock {\em Journal of the Royal Statistical Society: Series B (Statistical
  Methodology)}, 61(3):611--622, 1999.

\bibitem{hooram2015poisson}
Nam Hooram.
\newblock Poisson extension of gaussian process factor analysis for modeling
  spiking neural populations master’s thesis.
\newblock {\em Department of Neural Computation and Behaviour, Max Planck
  Institute for Biological Cybernetics, Tubingen}, 8, 2015.

\bibitem{lakshminarasimhan2018dynamic}
Kaushik~J Lakshminarasimhan, Marina Petsalis, Hyeshin Park, Gregory~C
  DeAngelis, Xaq Pitkow, and Dora~E Angelaki.
\newblock A dynamic bayesian observer model reveals origins of bias in visual
  path integration.
\newblock {\em Neuron}, 99(1):194--206, 2018.

\bibitem{balzani2020efficient}
Edoardo Balzani, Kaushik Lakshminarasimhan, Dora Angelaki, and Cristina Savin.
\newblock Efficient estimation of neural tuning during naturalistic behavior.
\newblock {\em Advances in Neural Information Processing Systems},
  33:12604--12614, 2020.

\bibitem{wilson2013gaussian}
Andrew Wilson and Ryan Adams.
\newblock Gaussian process kernels for pattern discovery and extrapolation.
\newblock In {\em International conference on machine learning}, pages
  1067--1075. PMLR, 2013.

\bibitem{de2021mogptk}
Taco de~Wolff, Alejandro Cuevas, and Felipe Tobar.
\newblock Mogptk: The multi-output gaussian process toolkit.
\newblock {\em Neurocomputing}, 424:49--53, 2021.

\bibitem{rutten2020non}
Virginia Rutten, Alberto Bernacchia, Maneesh Sahani, and Guillaume Hennequin.
\newblock Non-reversible gaussian processes for identifying latent dynamical
  structure in neural data.
\newblock {\em Advances in neural information processing systems},
  33:9622--9632, 2020.

\bibitem{zhao2017variational}
Yuan Zhao and Il~Memming Park.
\newblock Variational latent gaussian process for recovering single-trial
  dynamics from population spike trains.
\newblock {\em Neural computation}, 29(5):1293--1316, 2017.

\bibitem{nassar2018learning}
Josue Nassar, Scott~W Linderman, Yuan Zhao, M{\'o}nica Bugallo, and Il~Memming
  Park.
\newblock Learning structured neural dynamics from single trial population
  recording.
\newblock In {\em 2018 52nd Asilomar Conference on Signals, Systems, and
  Computers}, pages 666--670. IEEE, 2018.

\bibitem{duncker2019learning}
Lea Duncker, Gergo Bohner, Julien Boussard, and Maneesh Sahani.
\newblock Learning interpretable continuous-time models of latent stochastic
  dynamical systems.
\newblock In {\em International Conference on Machine Learning}, pages
  1726--1734. PMLR, 2019.

\bibitem{duncker2021dynamics}
Lea Duncker and Maneesh Sahani.
\newblock Dynamics on the manifold: Identifying computational dynamical
  activity from neural population recordings.
\newblock {\em Current opinion in neurobiology}, 70:163--170, 2021.

\end{thebibliography}


\begin{thebibliography}{1}

\bibitem{tipping1999probabilistic}
Michael~E Tipping and Christopher~M Bishop.
\newblock Probabilistic principal component analysis.
\newblock {\em Journal of the Royal Statistical Society: Series B (Statistical
  Methodology)}, 61(3):611--622, 1999.

\bibitem{bach2005probabilistic}
Francis~R Bach and Michael~I Jordan.
\newblock A probabilistic interpretation of canonical correlation analysis.
\newblock Technical report, 2005.

\bibitem{bishop2006pattern}
Christopher~M Bishop and Nasser~M Nasrabadi.
\newblock {\em Pattern recognition and machine learning}, volume~4.
\newblock Springer, 2006.

\bibitem{keeley2020}
Keeley S.L. Aoi M.C. Yu Y. Smith S.L.~Pillow J.W.
\newblock Identifying signal and noise structure in neural population activity
  with gaussian process factor models.
\newblock {\em NeurIPS}, 34, 2020.

\end{thebibliography}
\bibliographystyle{unsrt}

\end{document}


\maketitle
\beginsupplement

\section{Background on pPCA, pCCA, and their relation to TAME-GP}
\subsection{Canonical Correlation Analysis}
Given a random vector $\bm{x}$, PCA aims to find a linear transformation such that the components of the transformed vector are uncorrelated. In other words, it tries to find a linear transformation that diagonalizes the co-variance matrix of the random vectors. Similarly, CCA starts from two random vectors $\bm{x}_1$ and $\bm{x}_2$ of dimensions $m_1$ and $m_2$, and tries to find two linear transformations $U \in \mathbbm{R}^{m_1 \times m_1}$ and $V \in \mathbbm{R}^{m_2 \times m_2}$ such that each component of $U \cdot \bm{x}_1$ is correlated with a single component of $V \cdot \bm{x}_2$. In terms of correlation matrix, this corresponds to, 

\begin{equation}
\text{corr}\left(U \cdot \bm{x}_1, V \cdot \bm{x}_2\right)_{ij} = \begin{cases}
\rho_{i} & \text{if} \;i = j\\
0 & \text{otherwise}
\end{cases}
\end{equation}

where $\rho_i$ are called canonical correlations. Letting the joint empirical co-variance be $\hat{\Sigma} = \begin{bmatrix}
 \hat{\Sigma}_{11} & \hat{\Sigma}_{12} \\
 \hat{\Sigma}_{21} & \hat{\Sigma}_{22}
\end{bmatrix},$

it turns out that CCA projections are the singular vectors of the correlation matrix re-scaled by the inverse square-root of the individual co-variances. Namely, if $\tilde{u}_i, \tilde{v}_i$ are the i-th singular vectors of the correlation matrix $\text{corr}\left( \bm{x}_1,  \bm{x}_2\right)=\hat{\Sigma}_{11}^{-1/2} \hat{\Sigma}_{12} \hat{\Sigma}_{22}^{-1/2}$, then the canonical vectors are $(U_i,V_i) = (\hat{\Sigma}_{11}^{-1/2} \tilde{u}_i, \hat{\Sigma}_{22}^{-1/2} \tilde{v}_i)$.
The two projection matrices $U$ and $V$ are obtained by stacking the canonical vectors; it is immediate to verify that $I_{m_1} = U\tr \hat{\Sigma}_{11} U$, $ I_{m_2}= V\tr \hat{\Sigma}_{22} V$ and $P = U\tr \hat{\Sigma}_{12} V$, where $P$ is an $m_1 \times m_2$ diagonal matrix with diagonal entries the canonical correlations.\newline

Again, making the parallel with PCA, we know that the first PCA vector is the eigenvector of the empirical co-variance corresponding to the largest eigenvalue and satisfies  $\bm{w}_1 = \underset{ \Vert \bm{w} \Vert = 1}{\text{argmax}} \; \bm{w}\tr \text{cov}(\bm{x}) \bm{w}$. Similarly, it can be shown that the canonical vector corresponding to the largest singular value of the correlation matrix satisfies,

\begin{equation}
    (U_1, V_1) = \underset{\Vert \bm{u} \Vert = 1, \Vert \bm{v}\Vert=1}{\text{argmax}} \;\text{corr}(\bm{u}\tr \cdot \bm{x}_1, \bm{v}\tr \cdot \bm{x}_2).
\end{equation}
Finally the n-th canonical vector satisfies,
\begin{equation}
    (U_n, V_n) = \underset{\bm{u}\in \mathcal{U}^\perp, \bm{v}\in \mathcal{V}^\perp}{\text{argmax}} \;\text{corr}(\bm{u}\tr \cdot \bm{x}_1, \bm{v}\tr \cdot \bm{x}_2).
\end{equation}

with $\mathcal{U}^\perp = \{\bm{u}:\, \Vert \bm{u}\Vert = 1, \; \bm{u} \in \langle U_{1},\cdots U_{n-1} \rangle^\perp\}$ and $\mathcal{V}^\perp = \{\bm{v}:\, \Vert \bm{v}\Vert = 1, \; \bm{v} \in \langle V_{1},\cdots V_{n-1}\rangle^\perp\}$.\newline

\subsection{The probabilistic interpretation of PCA and CCA}
As first shown by Tipping and Bishop \cite{tipping1999probabilistic}, PCA can be expressed in terms of the maximum likelihood solution of the following probabilistic latent variable model,

\begin{align}
    p(\bm{z}) &\sim \mathcal{N} (0,I)\label{enq:pPCA:prior}\\
    p(\bm{x} | \bm{z}) &\sim \mathcal{N} (W \bm{z} +\mu,I,\sigma^2 I)\label{enq:pPCA:cond},
\end{align}
where $I$ is the $D \times D$ identity matrix, $W$ is a $N \times D$ projection matrix, $\mu \in \mathbb{R}^N$ is an intercept term and $\sigma^2$ a positive constant. 

Similarly,  Bach and Jordan \cite{bach2005probabilistic} showed that the canonical directions emerge from the maximum likelihood estimates of a simple probabilistic model,

\begin{alignat}{2}
    \bm{z} &\sim \mathcal{N}(0,I_D) \qquad && D = \text{min}(m_1,m_2)\label{enq:prior-pCCA}\\
    \bm{x_1} | \bm{z} &\sim \mathcal{N}(W_1\bm{z} + \mu_1, \Psi_1) \qquad && \Psi_1 \succeq  0\label{enq:cond1-pCCA}\\
    \bm{x_2} | \bm{z} &\sim \mathcal{N}(W_2\bm{z} + \mu_2, \Psi_2) \qquad && \Psi_2 \succeq  0\label{enq:cond2-pCCA}.
\end{alignat}

where we use a notation similar to that of equations (\ref{enq:pPCA:prior}, \ref{enq:pPCA:cond}) for the projection weights, the intercept and the identity matrix, while $\Psi_1$ and $\Psi_2$ are generic positive semi-definite $N \times N$ matrices.

We will refer to these models as the probabilistic PCA and probabilistic CCA, or pPCA and pCCA.

Too better highlight the link between CCA and pCCA we report the ML estimates of the pCCA model parameters, 

\begin{alignat}{2}
    \hat{W}_1 &= \hat{\Sigma}_{11} U M_1 \\
    \hat{W}_2 &= \hat{\Sigma}_{22} V M_1 \\
    \hat{\Psi}_1 &= \hat{\Sigma}_{11} - \hat{W}_1 \hat{W}_1 \tr\\
    \hat{\Psi}_2 &= \hat{\Sigma}_{22} - \hat{W}_2 \hat{W}_2 \tr \\
    \hat{\mu}_1 &= \frac{1}{N}\sum_j x_{1j}\\
    \hat{\mu}_2 &= \frac{1}{N}\sum_j x_{2j},
\end{alignat}

where $M_i$ are arbitrary $D \times D$ matrices such that $M_1 M_2\tr = P$, the diagonal matrix of the canonical correlations, $U$ and $V$ are the canonical directions.\newline

The posterior means and co-variances are given by,

\begin{alignat}{2}
    \mathbbm{E}[\bm{z}|\bm{x}_1] &= M_1\tr U\tr (\bm{x}_1-\hat{\mu}_1) \\
    \mathbbm{E}[\bm{z}|\bm{x}_2] &=  M_2\tr V\tr (\bm{x}_2-\hat{\mu}_2)\\
    \text{cov}\left(\bm{z}|\bm{x}_1\right) &= I - M_1 M_1\tr \\
    \text{cov}\left(\bm{z}|\bm{x}_2\right) &= I - M_2 M_2\tr\\
    \mathbbm{E}[\bm{z}|\bm{x}_1, \bm{x}_2] &= \begin{bmatrix}
        M_1 \\
        M_2
    \end{bmatrix} 
    \begin{bmatrix}
        (I - P^2)^{-1}   &  (I - P^2)^{-1} P \\
        (I - P^2)^{-1} P & (I - P^2)^{-1}
    \end{bmatrix}
    \begin{bmatrix}
        U\tr (\bm{x}_1-\hat{\mu}_1) \\
        V\tr (\bm{x}_2-\hat{\mu}_2)
    \end{bmatrix}\\
    \text{cov}\left(\bm{z}|\bm{x}_1,\bm{x}_2\right) &= I - \begin{bmatrix}
        M_1 \\
        M_2
    \end{bmatrix} 
    \begin{bmatrix}
        (I - P^2)^{-1}   &  (I - P^2)^{-1} P \\
        (I - P^2)^{-1} P & (I - P^2)^{-1}
    \end{bmatrix}
    \begin{bmatrix}
        M_1 \\
        M_2
    \end{bmatrix} \tr.
\end{alignat}

It is important to notice that, independently of the $M_1$ and $M_2$ matrices, the observation gets projected into the $D$-dimensional subspace of the canonical directions. See \cite{bishop2006pattern} for a similar argument bridging PCA and pPCA. 

\subsection{TAME-GP combines and extends the pPCA and pCCA generative models}

The probabilistic interpretation of PCA and CCA, Eqs. (\ref{enq:pPCA:prior}-\ref{enq:cond2-pCCA}) - allows (1) extending the model to non-Gaussian observation noise, (2) replacing the normal prior over the latent with a smoothing GP-prior, and (3) combining the two graphical models in a more general framework.

In particular, TAME-GP assumes a shared latent factor $\bm{z}^{(0)}$ with a GP prior that captures fine time scale correlations between some continuous task variables of interest (modelled as conditionally Gaussian) and the spike counts from multiple brain regions (modelled as conditionally Poisson). This approach extends the ideas of pCCA to the analysis of spike trains driven by smooth temporal dynamics. Further, we extended our graphical model by including additional area-specific latent factors $\bm{z}^{(j)}$ (GP-distributed). The projection associated with those factors aim specifically to capture the residual inter-area co-fluctuations, in close resemblance to the role of the pPCA projection weights. 

The general formulation of the TAME-GP generative model is given by Eqs.1-3 in the main text.

\section{Inverting the Hessian of the joint log-likelihood}
The dimensionality of the individual latents and trial duration pose computational challenges for TAME-GP approximate inference. For each trial, evaluating the posterior covariance requires inverting the Hessian of the joint log-likelihood, of dimensionality  $D\times D$, where $D = T \sum_j d_j$, $d_j$ is the dimension of $\bm{z}^{(j)}$ and $T$ is the number of time points of the trial (for simplicity, we assume all trials are the same length here, but the implementation allows for variability in trial duration). Hence, a naive implementation of the posterior estimation would require $O\left(D^3 \right)$ operations (the cost of inverting a $D$-dimensional matrix). Nonetheless,  the specific conditional independence assumptions of our model allow us to speed up this computation by using the block matrix inversion theorem. In particular, if we define
\begin{align*}
    \nabla_{\bm{z}^{(h)}} \nabla_{\bm{z}^{(k)}} \log p (\bm{z},\bm{x},\bm{y}) \equiv H_{h k},
\end{align*}
$\bm{H}$ has the following structure,
\begin{align*}
    H = 
    \begin{bmatrix}
        H_{00}    & H_{01} & H_{02}  & \cdots  & H_{0n}\\
        H_{01}\tr & H_{11} & \bm{0}  & \cdots  & \bm{0}\\
        H_{02}\tr & \bm{0} & H_{22}  & \cdots  & \bm{0}\\
                  &        &    \ddots      &  &  \\
        H_{0n}\tr & \bm{0}& \bm{0}  & \cdots  & H_{nn}
    \end{bmatrix},
\end{align*}
therefore, it can be inverted according to,
\begin{align*}
    \begin{bmatrix}
    A & C\tr \\
    C & B
    \end{bmatrix}^{-1} = \begin{bmatrix}
    (A- C\tr B^{-1} C)^{-1} & -(A-C\tr B^{-1} C)^{-1} C\tr B^{-1}\\
    - C B^{-1} (A-C\tr B^{-1} C)^{-1} &  B^{-1} + B^{-1}C(A-C\tr B^{-1}C)C\tr B^{-1}
    \end{bmatrix},
\end{align*}
by setting $A = H_{00}$ and $B=\begin{bmatrix}
 H_{11} & \bm{0} & \cdots & \bm{0}\\
\bm{0} &  H_{22}& \cdots & \bm{0}\\
      &        & \ddots & \\
\bm{0} & \bm{0} & \cdots & H_{nn}
\end{bmatrix}$, and $C = \begin{bmatrix} H_{01}\tr\\ \vdots \\ H_{0n}\tr \end{bmatrix}$; computing $B^{-1}$ requires only inverting the block-diagonal elements, while  $(A - C\tr B^{-1} C)$ has the same size as $H_{00}$, achieving an inversion of $\bm{H}$ in $O(T^3\sum_j d_j^3 )$ operations.

\section{Learning the Poisson observation parameters}
In order to learn the Poisson observation parameters we numerically maximize $\mathbbm{E}_q \left[\log(p(\mathbf{x},\mathbf{y},\mathbf{z} |\boldsymbol{\theta})\right]$ as a function of  $W^{(0,j)}$,  $W^{(j,j)}$ and $\mathbf{h}^{(j)}$ \footnote{$\bm{\theta}=\{\mathbf{W}^{(0/j,j)},\mathbf{h}^{(j)},\mathbf{C},\mathbf{d},\boldsymbol{\Psi},\mathbf{\tau}^{(j)}\}$}. Our implementation follows a Newton scheme which requires both the gradient and the Hessian of the optimization objective.

In order to simplify notation, we fix a unit $i$ from population $j$ and we set 
\begin{align*}
\bm{\mu}_t &= \begin{bmatrix}
    \bm{\mu}^{(0)}_t\\
    \bm{\mu}^{(j)}_t
\end{bmatrix}\\
\Sigma_t &= \begin{bmatrix}
    \Sigma^{(0,0)}_t && \Sigma^{(0,j)}_t\\
    \Sigma^{(0,j)\top}_t && \Sigma^{(j,j)}_t
\end{bmatrix}\\
W &= \begin{bmatrix}
    W_i^{(0,j)\;\top}\\
    W_i^{(j,j)\;\top}
\end{bmatrix}\\
x_t &= x^{(j)}_{it}\\
h&=h_i^{(j)},
\end{align*}
where $W\in\mathbb{R}^{d_0+d_j}$, and $h\in\mathbb{R}$.
The corresponding gradient and derivative will be,
\begin{align}
    \frac{\partial \mathbbm{E}_q \left[\log(p(\mathbf{x},\mathbf{y},\mathbf{z} |\boldsymbol{\theta})\right]}{\partial W} &= \sum_{l,t} x_t \bm{\mu}_t - \text{e}^{h+W\tr\bm{\mu}_t+\frac{1}{2}W\tr\Sigma_t W} \left( \mu_t +  \Sigma_t W\right)\\
    \frac{\partial \mathbbm{E}_q \left[\log(p(\mathbf{x},\mathbf{y},\mathbf{z} |\boldsymbol{\theta})\right]}{\partial h} &= \sum_{l,t}  x_t - \text{e}^{h+W\tr\bm{\mu}_t+\frac{1}{2}W\tr\Sigma_t W} \\
    \frac{\partial^2 \mathbbm{E}_q \left[\log(p(\mathbf{x},\mathbf{y},\mathbf{z} |\boldsymbol{\theta})\right]}{\partial W^2} &= -  \text{e}^{h+W\tr\bm{\mu}_t+\frac{1}{2}W\tr\Sigma_t W}\left[ \left(\mu_t +  \Sigma_t W\right) \left( \mu_t +  \Sigma_t W\right)\tr + \Sigma_t \right]\\
    \frac{\partial^2 \mathbbm{E}_q \left[\log(p(\mathbf{x},\mathbf{y},\mathbf{z} |\boldsymbol{\theta})\right]}{\partial h \partial W} &= -  \text{e}^{h+W\tr\bm{\mu}_t+\frac{1}{2}W\tr\Sigma_t W} \left( \mu_t +  \Sigma_t W\right)\\
    \frac{\partial^2 \mathbbm{E}_q \left[\log(p(\mathbf{x},\mathbf{y},\mathbf{z} |\boldsymbol{\theta})\right]}{\partial h^2 } &= -  \text{e}^{h+W\tr\bm{\mu}_t+\frac{1}{2}W\tr\Sigma_t W}
\end{align}

where $l=1,\dots,M$ and $t=1,\dots,T$ are the trial and time indexes respectively.

\section{Learning the GP time constants}
GP hyperparameters (time constant) are learned by gradient based numerical optimization of the joint log-likelihood. 
Following the notation of the main text we set, $\lambda^{(j)}_i  = -\log (2 \tau^{(j)}_i)$, and we define a kernel $\mathbf{K}^{(j)}_{i} : \mathbb{R} \longrightarrow \mathbb{R}^{T \times T}$ such that, $\left[\mathbf{K}^{(j)}_{i}(\lambda)\right]_{ts} = \exp \left( - e^{\lambda} (t-s)^2 \right)$.

The objective function takes the form,
$$
\mathbbm{E}_q \left[\log(p(\mathbf{x},\mathbf{y},\mathbf{z} |\boldsymbol{\theta}) \right] = \sum_{l,j,i}  -\text{trace} \left(\bm{K}^{(j) -1}_i (\lambda_i^{(j)}) \mathbb{E}_q[\bm{z}^{(j)}_i \bm{z}^{(j)\top}_i] \right) - \log |\bm{K}^{(j)}_i(\lambda_i^{(j)})| +\text{const},
$$
where $j=0,...,n$ is the latent factor, $l=1,...,M$ is the trial number and $i=1,...,d_j$ is the component of $\bm{z}^{(j)}$.  Using the chain rule we obtain,

$$
\frac{\partial \mathbbm{E}_q \left[\log(p(\mathbf{x},\mathbf{y},\mathbf{z} |\boldsymbol{\theta}) \right]}{\partial {\lambda}^{(j)}_i} =  \text{trace} \left( \frac{\partial \mathbbm{E}_q \left[\log(p(\mathbf{x},\mathbf{y},\mathbf{z} |\boldsymbol{\theta}) \right]}{\partial \bm{K}^{(j)}_i}\tr \cdot \frac{\partial \bm{K}^{(j)}_i}{\partial \lambda^{(j)}_i}   \right),
$$
with
\begin{align*}
    \frac{\partial \mathbbm{E}_q \left[\log(p(\mathbf{x},\mathbf{y},\mathbf{z} |\boldsymbol{\theta}) \right]}{\partial \bm{K}^{(j)}_i} &= \frac{1}{2}  \sum_l \left( -K^{(j) -1}_i + K^{(j) -1}_i \mathbb{E}_q[\bm{z}^{(j)}_i \bm{z}^{(j)\top}_i] K^{(j) -1}_i\right)\\
    \frac{\partial \left[K^{(j)}_i\right]_{t s}}{\partial \lambda} &=  -\text{e}^{\lambda} (t-h)^2 \exp\left(-\text{e}^{\lambda} (t-s)^2\right).
\end{align*}

\section{Parameter initialization}
\textbf{Factorized TAME}. Before running EM on the full TAME, we obtain initial condition for the model parameters (all except the GP kernel hyperparameters) by means of running five iterations of EM for the temporally factorized version of the model. In particular, we replace the GP-prior over the latents with a product of a Gaussian normal distributions, i.e. $p(\bm{z}^{(j)}_i )= \prod_t p(z^{(j)}_{it})$, and $p(z^{(j)}_{it})\sim \mathcal{N}(0,1)$. 

Under this prior assumption the joint likelihood as a whole factorizes over the temporal axis (i.e. the observations are temporally independent given the latents).
As a consequence, the Hessian matrix of the joint pdf is sparse, and can be stored and  inverted efficiently, allowing for the implementation of a full Newton scheme to numerically optimize for the MAP estimate of the posterior ever latents $\bm{z}$.

The EM-based optimization of the factorized TAME also needs an initial choice for parameters. We found empirically that a CCA-based heuristic works well for this purpose. Specifically, we set:
\begin{itemize}
    \item $W^{(0,j)}$ to the first $d_0$ canonical directions $V$ between the square-rooted, mean-centered spike counts of population $j$, $\bm{s}^{(j)} = \sqrt{\bm{x}^{(j)}} - \mu_j$ and the task variables $\bm{y}$ ($\mu_j$ is the empirical mean of the square-rooted spikes).
    \item $W^{(j,j)}$ as the first $d_j$ principal direction for the orthogonal complement of the counts w.r.t the canonical directions,
$\bm{s}^{(j)}_{\text{ort }t} = \bm{s}^{(j)}_t - V\tr V \bm{s}^{(j)}_t.$ This will initially enforce orthogonality in the task relevant and private latent subspaces.
\item $\bm{h}^{(j)}$ was set to the log of the empirical mean of the counts.
\item $C$ was set to the first $d_0$ canonical directions $U$ between $\bm{s}$ and the square-rooted counts from all the neural populations, $\bm{Y} = [\bm{y}^{(1)}; \dots ;\bm{y}^{(m)}]$.
\item $\bm{d}$ was set to the empirical mean of $\bm{s}$, and $\bm{\Psi}$ to the empirical covariance.
\end{itemize}

\textbf{GP time constants.} The initial GP time constants were drawn from a uniform random distribution $\tau^{(j)}_i \sim \text{U}[0,0.5]$.

\section{Comparison of TAME-GP and SNP-GPFA}
\label{sess:snp-gpfa}

We compared our framework to that of SNP-GPFA \cite{keeley2020}, which identifies shared fluctuation between multiple neural populations under the assumption of trial repeats with a common stimulus-driven mean (corresponding to a dimensionality reduced peristimulus time histogram, or PSTH). 

Briefly, the multi-area SNP-GPFA assumes that the spike counts of two areas, area A and area B, are generated according to

\begin{align}
    \begin{bmatrix}
    \bm{Y}^A_j \\
    \bm{Y}^B_j
    \end{bmatrix} = \text{Poisson} \left( f \left( \bm{W}_s \bm{X}^s +
    \begin{bmatrix}
    \bm{W}_{AA} & \bm{0} \\
    \\
    \bm{0} & \bm{W}_{BB}
    \end{bmatrix} \begin{bmatrix} \bm{X}^{A,n}_{j} \\ \\ \bm{X}^{B,n}_{j}
    \end{bmatrix}\right) \right),\label{eqn-snpgpfa}
\end{align}

with $\bm{Y}^{A/B}_j$ the spike counts of population $A$ and $B$ for trial $j$, $f$ the soft-max non-linearity, $\bm{X}^{A/B,n}_{j}$ a gaussian process (with factorized RBF covariance) capturing within area co-fluctuations for trial $j$,  $\bm{X}^{A/B,s}$ another gaussian process shared across-trial and populations capturing the between area co-fluctuations.

We generated spike counts from the graphical model in figure \ref{fig:snp-gpfa}A assuming a fixed trial duration (necessary for the SNP-GPFA), in different conditions: 1) fixing the shared dynamics across trials (as in SNP-GPFA, figure \ref{fig:snp-gpfa}B, top), or 2) varying the shared dynamics across trial (figure \ref{fig:snp-gpfa}B, bottom).

Specifically, for the first condition the counts followed (\ref{eqn-snpgpfa}), but replacing the non-linearity with an exponential. For the second case, the counts follow Poisson statistics of the form

\begin{align}
    \begin{bmatrix}
    \bm{Y}^A_j \\
    \bm{Y}^B_j
    \end{bmatrix} &= \text{Poisson} \left( \text{exp} \left( \bm{W}_s \bm{X}^s_{j} +
    \begin{bmatrix}
    \bm{W}_{AA} & \bm{0} \\
    \\
    \bm{0} & \bm{W}_{BB}
    \end{bmatrix} \begin{bmatrix} \bm{X}^{A,n}_{j} \\ \\ \bm{X}^{B,n}_{j}
    \end{bmatrix}\right) \right),
\end{align}

where we added the trial dependency to the shared Gaussian process factor.

We set the dimensionality of the shared factor to 2, and of each private factors to 3. We simulated spike counts from two populations of 30 neurons for 50 trials, each having 100 time points with a 0.05 second resolution. The average firing rate of both population was set to 10Hz. We fit the simulated spike counts with TAME-GP and SNP-GPFA for both conditions. The results show that TAME-GP captures the between area co-fluctuation in both scenarios while SNP-GPFA fails when the shared dynamics varies between trials, as expected by the model assumptions (Fig.~\ref{fig:snp-gpfa}C,E).
We assessed the accuracy of the factorization of the spike-count variance by means of Lasso regression. In particular, we regressed the ground truth latents from the estimated latents of the different models, and quantified regression goodness-of-fit in terms of cross-validated $R^2$  (Fig.~\ref{fig:snp-gpfa}D,F). We quantified the contribution of each latent factor to the regression in terms of the magnitude the associated coefficients. Results (reported in Table \ref{tabl:lasso}) show  that 1) both models can factorize the variance when the shared dynamics are fixed across trials, with SNP-GPFA achieving a cleaner decomposition (expected given that it is a closer model of the true data generating process in this case); 2) TAME-GP achieves a near optimal factorization when the shared latents vary across trials (as assumed by its generative model), while SNP-GPFA is unable to find the appropriate decomposition.

\section{Selecting the number of private and shared dimensions in real data}

We select the number of private and shared dimensions to fit in real data by optimizing these hyperparameters via a grid search. A priori we set the maximum number of dimensions to be evaluated as the number of PCs needed to account for 80\% of the population variance (in this case, 5 dimensions). Fig.\ref{fig:hypsel} shows estimates of model fit quality as a function of the number of dimensions included in private and shared latents for the multi-area TAME-GP presented in Fig.4I-K. The results show a well-behaved cross-validated $R^2$ landscape, with optimal dimensionalities $(5,5)$.

\newpage

\section{Supplemental Figures}
\FloatBarrier

\begin{figure}[H]
\centering
\includegraphics[width=0.5\textwidth]{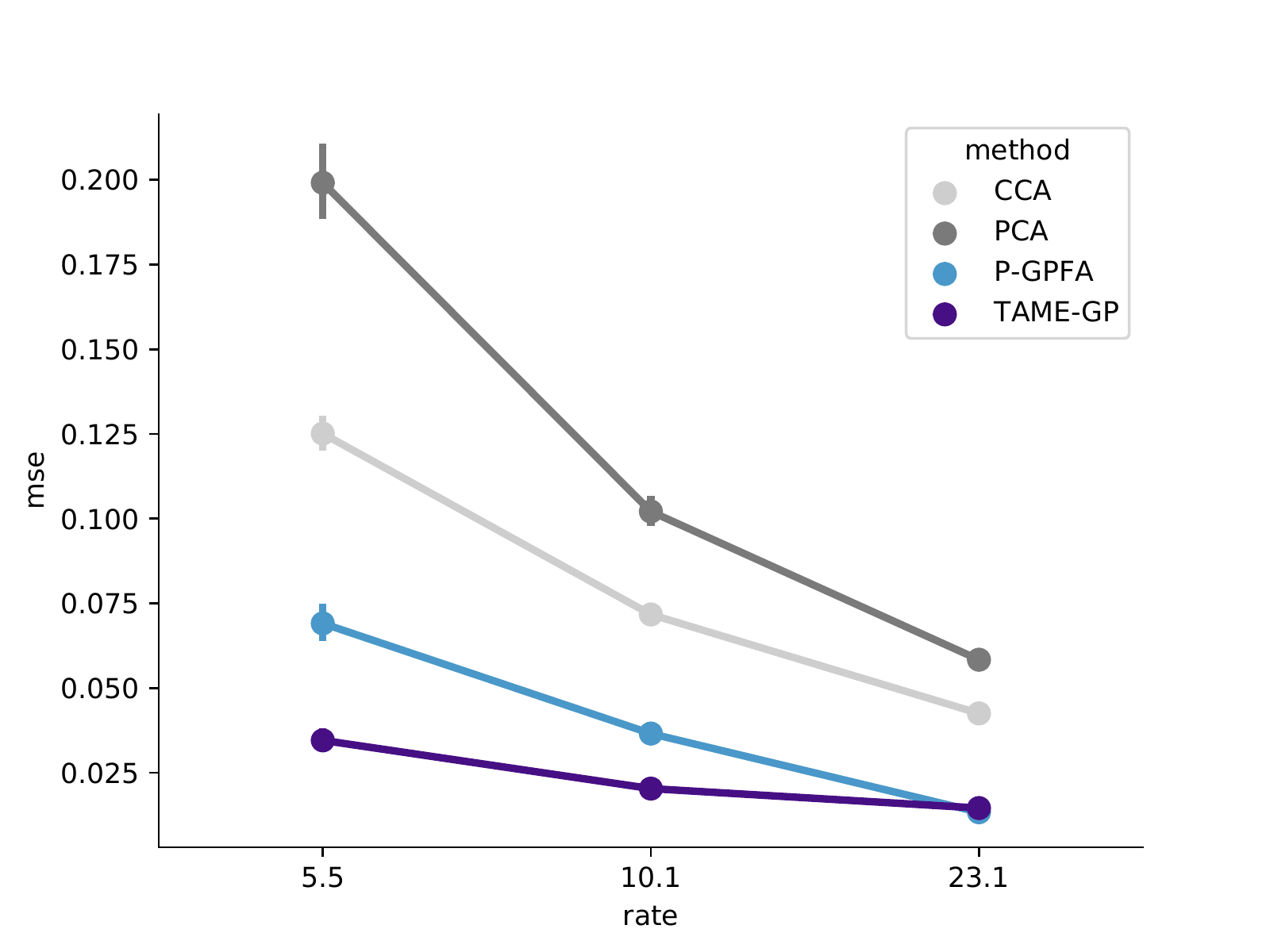}
\caption{Task-aligned latent dynamics reconstruction (extends fig. 2C). 
Mean squared error between the true task relevant dynamics and the model reconstruction based on the 2 dimensional task relevant latent factor for CCA and TAME-GP, and the full 6 dimensional latent space for P-GPFA and PCA. In contrast, figure 2C shows the MSE based only on the first 2 principal latents for pCCA and 
P-GPFA; Error bars represent the mean $\pm$ s.d. over 10-fold cross-validation.}
\label{fig:reconsError_monkey}
\end{figure}

\begin{figure}[h]
\centering
\includegraphics[width=0.9\textwidth]{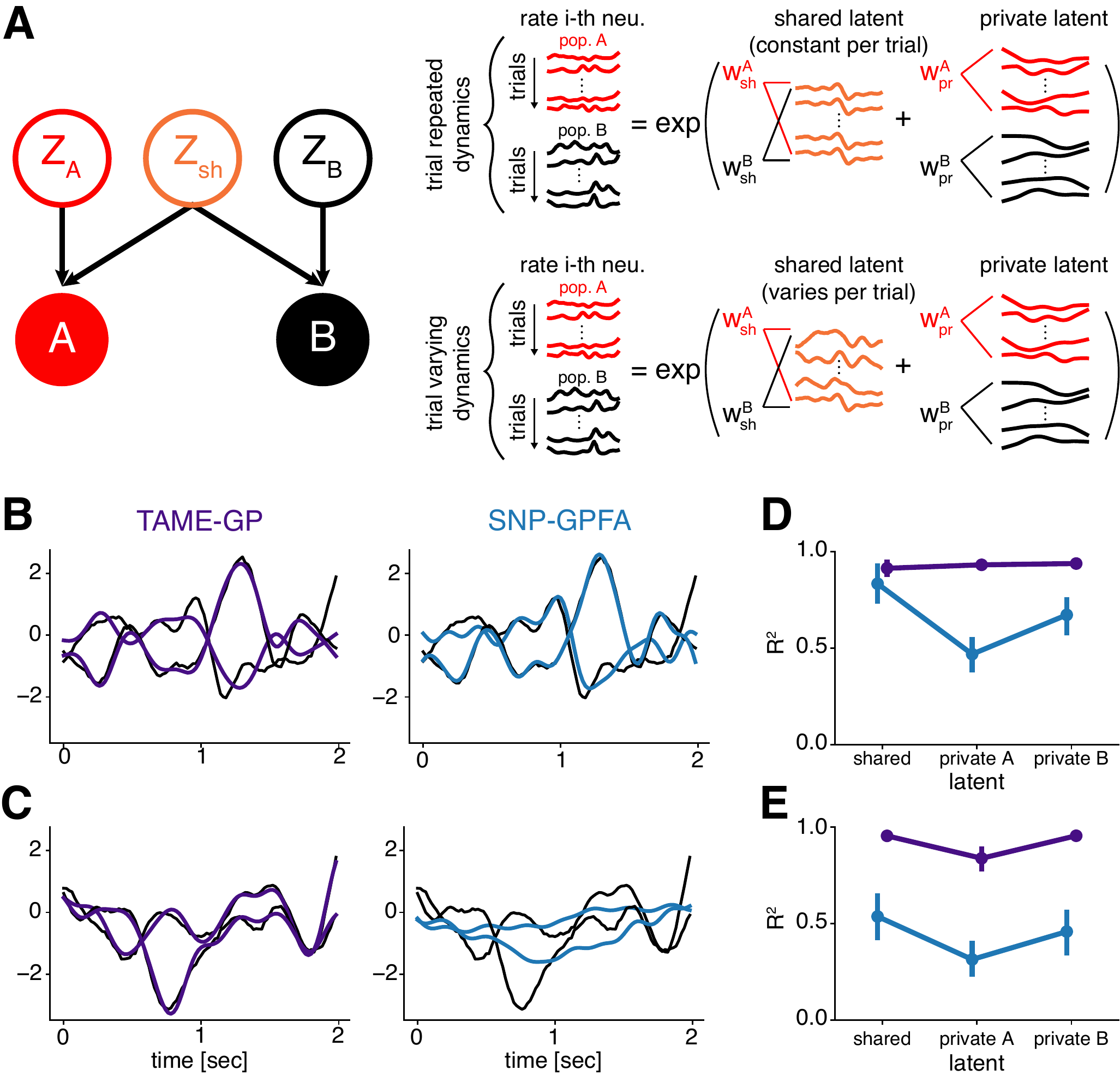}
\caption{Communication subspace estimation with SNP-GPFA and TAME-GP (extends figure 3). \textbf{A}. Scheme of the spike count generative model for trial repeated (top right) and trial varying (bottom right) shared dynamics. \textbf{B},\textbf{C}. Ground truth shared dynamics (black lines) and model reconstructions (colored lines) for the trial repeated (\textbf{B}) and trial varying (\textbf{C}) conditions. \textbf{D},\textbf{E}. Ground truth shared and private dynamics variance explained by model predictions for the trial repeated (\textbf{D}) and trial varying (\textbf{E}) conditions.; error bars represent mean $\pm$ standard deviation over a 5-fold cross validation. }
\label{fig:snp-gpfa}
\end{figure}

\begin{figure}[h]
\centering
\includegraphics[width=0.5\textwidth]{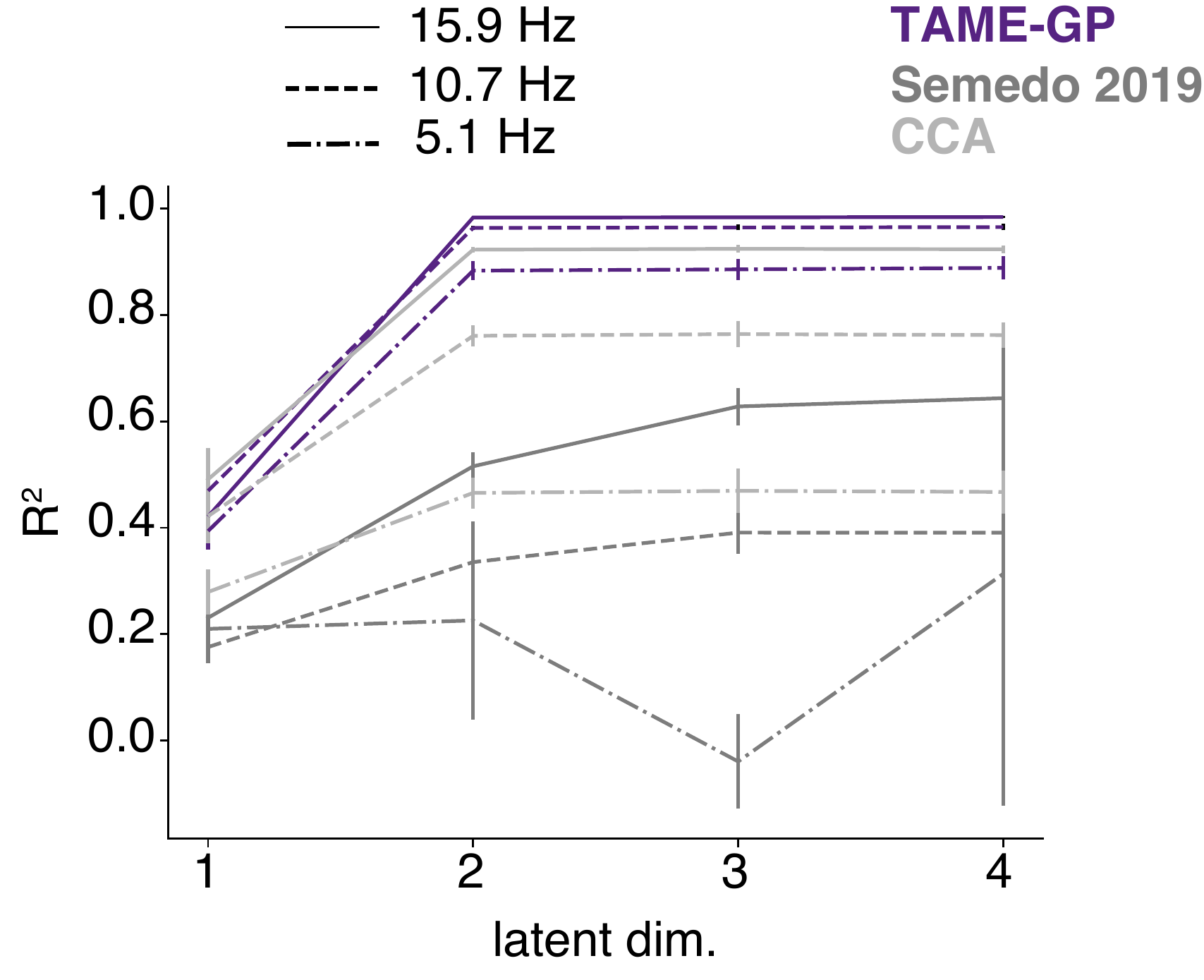}
\caption{ Model fit of shared and task aligned dynamics. R2 of the linear regression
between the ground truth task aligned latent dynamics and the model MAP estimate for TAME
(purple), PCCA (light grey) and reduced rank regression (dark grey). Extends fig. 3I in the main text to multiple average firing rates.}
\label{fig:reconsError}
\end{figure}

\begin{figure}[t]
\centering
\includegraphics[width=0.5\textwidth]{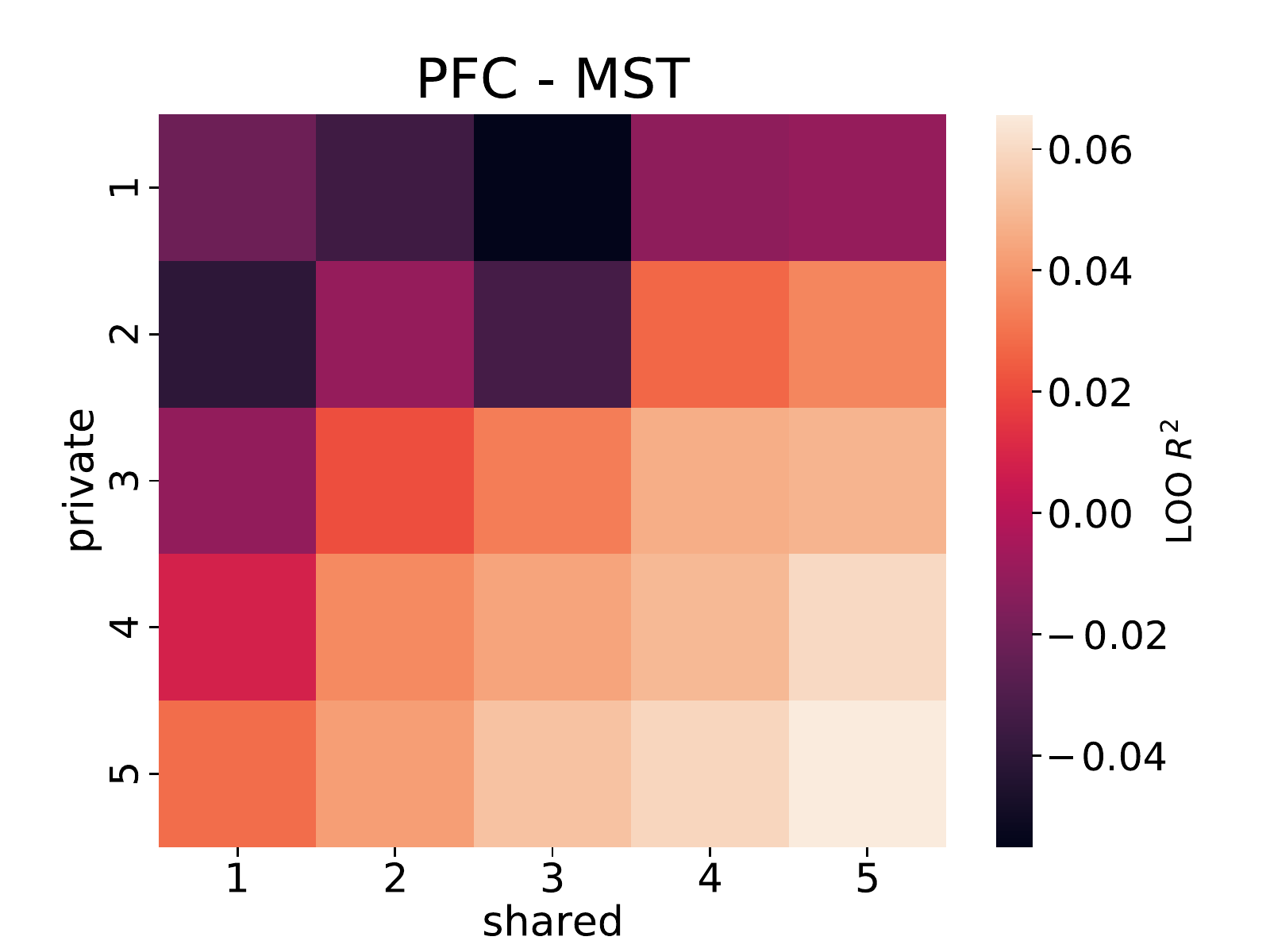}
\caption{Latent dimensionality selection, for the MSTd and dlPFC communication manifold analysis (extends fig. 4I-K). Heat-map of the leave-one-neuron-out R2 of the spike count variance explained by a TAME-GP for different combination of shared and private latent dimensions. The upper bound on dimensionality was set to the number of principal components needed to explain 80\% of the population spike count variance.}
\label{fig:hypsel}
\end{figure}


\FloatBarrier
\section{Supplementary table}
\begin{table}[H]
\centering
\begin{tabular}{|l|l|l|l|l|} 
\hline
\multicolumn{5}{|l|}{\textbf{Lasso results}}                                                                              \\ 
\hline
\textbf{model} & \textbf{sim type} & \textbf{ground truth latent} & \textbf{model latent} & $\Vert \mathbf{\beta} \Vert$  \\ 
\hline
SNP-GPFA       & fixed             & private A                    & private A             & 0.252458                      \\
               &                   &                              & private B             & 0.008377                      \\
               &                   &                              & shared                & 0.065025                      \\ 
\cline{3-5}
               &                   & private
  B                  & private A             & 0.001043                      \\
               &                   &                              & private B             & 0.373415                      \\
               &                   &                              & shared                & 0.012722                      \\ 
\cline{3-5}
               &                   & shared                       & private A             & 0.014438                      \\
               &                   &                              & private B             & 0.026413                      \\
               &                   &                              & shared                & 0.665547                      \\ 
\cline{2-5}
               & variable          & private A                    & private A             & 0.128838                      \\
               &                   &                              & private B             & 0.256987                      \\
               &                   &                              & shared                & 0.044492                      \\ 
\cline{3-5}
               &                   & private
  B                  & private A             & 0.046482                      \\
               &                   &                              & private B             & 0.332019                      \\
               &                   &                              & shared                & 0.114382                      \\ 
\cline{3-5}
               &                   & shared                       & private A             & 0.058657                      \\
               &                   &                              & private B             & 0.386729                      \\
               &                   &                              & shared                & 0.010308                      \\ 
\hline
TAME-GP        & fixed             & private A                    & private A             & 0.222231                      \\
               &                   &                              & private B             & 0.00472                       \\
               &                   &                              & shared                & 0.16166                       \\ 
\cline{3-5}
               &                   & private
  B                  & private A             & 0.016308                      \\
               &                   &                              & private B             & 0.419728                      \\
               &                   &                              & shared                & 0.02365                       \\ 
\cline{3-5}
               &                   & shared                       & private A             & 0.1016                        \\
               &                   &                              & private B             & 0.006841                      \\
               &                   &                              & shared                & 0.476519                      \\ 
\cline{2-5}
               & variable          & private A                    & private A             & 0.268177                      \\
               &                   &                              & private B             & 0.011153                      \\
               &                   &                              & shared                & 0.032323                      \\ 
\cline{3-5}
               &                   & private
  B                  & private A             & 0.003969                      \\
               &                   &                              & private B             & 0.411102                      \\
               &                   &                              & shared                & 0.020695                      \\ 
\cline{3-5}
               &                   & shared                       & private A             & 0.019493                      \\
               &                   &                              & private B             & 0.005826                      \\
               &                   &                              & shared                & 0.658865                      \\
\hline
\end{tabular}
\newline
\caption{Lasso regression coefficients, related to session~\ref{sess:snp-gpfa}. Norm of the coefficients of the Lasso regression between the ground truth latent dynamics and the SNP-GPFA/ TAME-GP predicted latents. Lasso hyperparameters are set by grid search with a 5-fold cross-validation procedure.}\label{tabl:lasso}
\end{table}

\FloatBarrier

\small
\bibliography{neurips2022_refs}
\bibliographystyle{unsrt}